\documentclass[twocolumn,aps,prx,superscriptaddress,floatfix,showpacs,notitlepage]{revtex4-1}
\usepackage[T1]{fontenc}
\usepackage[utf8]{inputenc}
\usepackage{amsmath}
\usepackage{amssymb}
\usepackage{cancel}
\usepackage{graphicx}

\makeatletter
\usepackage{bm}
\usepackage{amsthm}
\usepackage{amsfonts}\usepackage{latexsym}
\usepackage[margin,index]{fixme}
\usepackage[usenames,dvipsnames]{color}

\usepackage{amsfonts}
\usepackage{bbold}

\usepackage{braket}

\makeatother

\begin{document}

\title{Nanoscale `Dark State' Optical Potentials for Cold Atoms}

\author{M. Łącki}

\affiliation{Institute for Quantum Optics and Quantum Information of the Austrian
Academy of Sciences, A-6020 Innsbruck, Austria}

\affiliation{Institute for Theoretical Physics, University of Innsbruck, A-6020
Innsbruck, Austria}

\author{M. A. Baranov}

\affiliation{Institute for Quantum Optics and Quantum Information of the Austrian
Academy of Sciences, A-6020 Innsbruck, Austria}

\affiliation{Institute for Theoretical Physics, University of Innsbruck, A-6020
Innsbruck, Austria}

\author{H. Pichler}

\affiliation{Institute for Quantum Optics and Quantum Information of the Austrian
Academy of Sciences, A-6020 Innsbruck, Austria}

\affiliation{Institute for Theoretical Physics, University of Innsbruck, A-6020
Innsbruck, Austria}

\affiliation{ITAMP, Harvard-Smithsonian Center for Astrophysics, 60 Garden Street,
Cambridge, MA 02138, USA}

\affiliation{Physics Department, Harvard University, 17 Oxford Street, Cambridge,
Massachusetts 02138, USA}

\author{P. Zoller}

\affiliation{Institute for Quantum Optics and Quantum Information of the Austrian
Academy of Sciences, A-6020 Innsbruck, Austria}

\affiliation{Institute for Theoretical Physics, University of Innsbruck, A-6020
Innsbruck, Austria}
\begin{abstract}
We discuss generation of subwavelength optical barriers on the scale
of tens of nanometers, as conservative optical potentials for cold
atoms. These arise from nonadiabatic corrections to Born-Oppenheimer
potentials from dressed `dark states' in atomic $\Lambda$-configurations.
We illustrate the concepts with a double layer potential for atoms
obtained from inserting an optical subwavelength barrier into a well
generated by an off-resonant optical lattice, and discuss bound states
of pairs of atoms interacting via magnetic dipolar interactions. The
subwavelength optical barriers represent an optical `Kronig-Penney'
potential. We present a detailed study of the bandstructure in optical
`Kronig-Penney' potentials, including decoherence from spontaneous
emission and atom loss to open `bright' channels. 
\end{abstract}

\pacs{37.10.Jk,32.80.Qk,37.10.Vz}

\date{\today}

\maketitle
Optical potentials generated by laser light are a fundamental tool
to manipulate the motion of cold atoms with both conservative and
dissipative forces \cite{Cohen-Tannoudji2011,Lewenstein2012}. Paradigmatic
examples of conservative optical potentials are optical dipole traps
from a focused far off-resonant light beam, or optical lattices (OL)
generated by an off-resonant standing laser wave, as basis of the
ongoing experimental effort to realize atomic Hubbard models \cite{bloch2008many}.
The underlying physical mechanism is the second-order AC Stark shift
of an electronic atomic level, which is proportional to the light
intensity. Optical potential landscapes, which can be designed, will
thus reflect, and be limited by the achievable spatial variation of
the light intensity. For light in the far-field, i.e.~for optical
trapping far away from surfaces (compare \cite{Rauschenbeutel2014,Thompson2013,Kimble2015,Lukin2009,Lukin2012}),
this spatial resolution will thus be given essentially by the wavelength
of the light $\lambda$. In the quest to realize free-space optical
subwavelength structures for atoms \cite{yi2008state,Zubairy2008,nascimbene2015dynamic,Ritt:2006cb,brezger1999polarization,Salger2007,lundblad2008atoms}
we will describe and study below a family of conservative optical
potentials, which arise as \emph{nonadiabatic corrections to dark
states} (DSs) in atomic $\Lambda$-type configurations \cite{lukin2003colloquium,vitanov2016stimulated},
building on the strong nonlinear atomic response to the driving lasers.
The present scheme should allow the realization of \emph{optical barriers}
for atoms on the scale of tens of nanometers, and in combination with
traditional optical potentials and lattices the formation of a complex
`nanoscale' optical landscape for atoms. Our discussion should be
of particular interest for realizing many-atom quantum dynamics as
a strongly interacting many-body systems, where atomic energy scales
and interactions, such as magnetic dipolar couplings \cite{Ferlaino2016,Lev2016,Kadau:2016cb,FerrierBarbut:2016jo,dePaz:2013ff},
are strongly enhanced by subwavelength distances.

\begin{figure}
\includegraphics[clip,width=8.6cm]{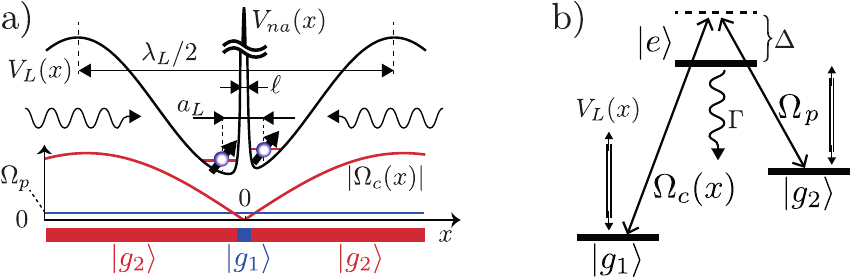} \caption{(Color online) (a) A double well potential for atoms is created by
inserting an optical subwavelength barrier $V_{na}(x)$ with width
$\ell$ into a potential well generated with an off-resonant OL $V_{L}(x)$
with lattice period $\lambda_{L}/2$, and size of the vibrational
ground state $a_{L}$, such that $\ell\ll a_{L}\ll\lambda_{L}/2$.
The subwavelength barrier is obtained with an atomic $\Lambda$-system
supporting a `dark state' as superposition of the two atomic ground
states $\ket{g_{1}}$ and $\ket{g_{2}}$ (b), where a resonant Raman
coupling from a strong control field $\Omega_{c}(x)=\Omega_{c}\sin(kx)$
($k=2\pi/\lambda$) and a weak probe field $\Omega_{p}$ connects
the two ground states (see text). }
\end{figure}

To illustrate the `nanoscale' optical potentials we can construct,
we show in Fig.~1 a setup, where a subwavelength barrier of width
$\ell$ is inserted into a potential well. This potential well can
be created, for example, with a (standard) off-resonant OL $V_{L}(x)=V_{0}\sin^{2}(k_{L}x)\approx V_{0}(k_{L}x)^{2}$
with $\lambda_{L}\equiv2\pi/k_{L}$ the wavelength of the trapping
laser, and we denote its ground state size by $a_{L}$. Thus our aim
is to create a double well potential for atoms on the subwavelength
scale $\ell\ll a_{L}\ll\lambda_{L}/2$. By adjusting the height, and
by displacing the subwavelength barrier we can control the tunnel
coupling between the wells, strongly enhanced relative to the standard
OL with lattice period $\lambda_{L}/2$. In a 3D (2D) setup this realizes
a \emph{double layer (wire)}, \emph{with subwavelength separation}.
Loading magnetic atoms or polar molecules with dipolar interactions
into these structures we benefit from the strongly enhanced energy
scales for interlayer(wire) interactions.

We propose and analyze below the physical realization of such a setup,
and we will mainly focus on a 1D model considering atomic motion along
$x$. The subwavelength barrier is obtained by choosing an atomic
$\Lambda$-transition with two long-lived ground (spin) states $\ket{g_{1}}\equiv\ket{\downarrow}$,
$\ket{g_{2}}\equiv\ket{\uparrow}$ (Fig.~1b) \cite{gardiner2015quantum,gorshkov2008coherent},
which are coupled by a Raman transition. The first leg of the Raman
coupling is a strong control laser with Rabi frequency $\Omega_{c}(x)=\Omega_{c}\sin(kx)$
as a standing wave with wavelength $\lambda=2\pi/k$ along $x$, and
the second is a weak probe laser with Rabi frequency $\Omega_{p}$
with propagation direction perpendicular to the axis $x$ \footnote{This is in contrast to spin-orbit coupling schemes based on running
wave laser configurations with $\Lambda$-systems \cite{Spielman2013}}. We denote the ratio of Rabi frequencies as $\epsilon\equiv\Omega_{p}/\Omega_{c}\ll1$.
The lasers are tuned to satisfy the Raman resonance condition, while
the detuning $\Delta$ from the excited state $\ket{e}$ can be near
or off-resonant. The relevant Hamiltonian is $H=-\hbar^{2}\partial_{x}^{2}/2m+H_{a}(x)$
\cite{gardiner2015quantum}, as a sum of the kinetic energy and the
internal atomic Hamiltonian 
\begin{align*}
H_{a} & \!=\hbar\!\left[\left(\!-\!\Delta\!-\!i\frac{\Gamma}{2}\!\right)\ket{e}\!\bra{e}\!+\!\frac{\Omega_{c}(x)}{2}\ket{e}\!\bra{g_{1}}\!+\!\frac{\Omega_{p}}{2}\ket{e}\!\bra{g_{2}}\!+\!{\rm h.c.}\!\right]
\end{align*}
written in a rotating frame and with $\Gamma$ the spontaneous decay
rate of $\ket{e}$. We can add to the above Hamiltonian a trapping
potential for the ground states $V(x)$ to generate the well of Fig.~1.
This is realized, e.g. as a 1D off-resonant lattice $V_{L}(x)=V_{0}\sin^{2}(k_{L}x)$
with an effective $k_{L}\!=\!2\pi/\lambda_{L}$, i.e. $V(x)\equiv V_{L}(x)\sum_{i=1,2}\ket{g_{i}}\bra{g_{i}}$.
This far off-resonant OL potential acts equally on both ground states,
and thus preserves the resonance Raman condition independent of $x$.

We are interested in the regime of slow atomic motion, where the kinetic
energy (and trapping potential $V(x)$) are small relative to the
energy scales set by $H_{a}$. In the spirit of the Born-Oppenheimer
(BO) approximation \cite{kazantsev1990mechanical,Olshanii1996,Dutta1999,Ruseckas2005}
we diagonalize $H_{a}(x)\ket{E_{\sigma}(x)}=E_{\sigma}(x)\ket{E_{\sigma}(x)}$
($\sigma=0,\pm$) to obtain position dependent dressed energies, 
\[
E_{0}(x)=0,\quad E_{\pm}(x)=\frac{\hbar}{2}\left[-\tilde{\Delta}\pm\sqrt{\Omega_{p}^{2}+\Omega_{c}^{2}(x)+\tilde{\Delta^{2}}}\right]
\]
with $\tilde{\Delta}\equiv\Delta+i\frac{\Gamma}{2}$, playing the
role of adiabatic BO potentials for the atomic motion. Such a $\Lambda$-configuration
supports an atomic DS $E_{0}=0$ as a linear combination of the ground
states, $\ket{E_{0}(x)}=-\cos\alpha(x)\ket{g_{1}}+\sin\alpha(x)\ket{g_{2}}$
with $\alpha(x)=\arctan[\Omega_{c}(x)/\Omega_{p}]$, which for an
atom at a given position $x$ (at rest) is decoupled from the exciting
Raman beams. The identity of this DS changes in space on a subwavelength
scale \cite{gorshkov2008coherent,Zubairy2008}: in regions $|\Omega_{c}(x)|\!\gg\!\Omega_{p}$
we have $\ket{E_{0}}\sim\ket{g_{2}}$, while $|\Omega_{c}(x)|\ll\Omega_{p}$
defines a region $\ell\!\equiv\!\epsilon\lambda/2\pi\ll\!\lambda$
with $\ket{E_{0}}\sim\ket{g_{1}}$ and thus a spatial \emph{subwavelength
spin structure} (bottom of Fig.~1a).

An atom prepared in the DS, and moving slowly in space will, in accordance
with an adiabaticity requirement \cite{bergmann2015perspective},
remain in this DS, and the internal state will change its internal
spin identity according to $\ket{E_{0}(x)}$ on the scale $\ell\!\ll\!\lambda$.
Correspondingly, there will be nonadiabatic corrections to this motion.
As shown below, these nonadiabatic corrections take on the form of
a \emph{subwavelength optical barrier} representing a \emph{conservative
potential} 
\begin{equation}
V_{na}(x)=\frac{\hbar^{2}}{2m}\left(\partial_{x}\alpha\right)^{2}\equiv E_{R}\frac{\epsilon^{2}\cos^{2}(kx)}{[\epsilon^{2}+\sin^{2}(kx)]^{2}}\label{eq:Vna}
\end{equation}
with $E_{R}=\hbar^{2}k^{2}/2m$ the recoil energy and atomic mass
$m$. The effective 1D Hamiltonian for the atomic motion is thus $h=-\hbar^{2}\partial_{x}^{2}/2m+V_{L}(x)+V_{na}(x)$.
In Fig.~1a this realizes the subwavelength barrier, where the vibrational
ground state of the OL potential $V_{L}(x)$ of size $a_{L}$ is split
into a double well for $\ell\ll a_{L}$. We note that $V_{na}(x)$,
apart from the overall scale $E_{R}$, depends only on $\epsilon=\Omega_{p}/\Omega_{c}$.
For $\epsilon\ll1$, $V_{na}(x)$ is a sequence of potential hills
with spacing $\lambda/2$, width $\ell\equiv\epsilon\lambda/2\pi\ll\lambda/2$
and height $E_{R}/\epsilon^{2}\gg E_{R}$ (c.f.~Fig.~2b), and has
for $\epsilon\ll1$ a form reminiscent of a repulsive Kronig-Penney
$\delta$-like comb $V_{na}(x)\rightarrow\sum_{n}E_{R}\lambda/(4\epsilon)\delta(x-n\lambda/2)$
\cite{Merzbacher1998}. For Raman beams derived from the same laser
source this potential is insensitive to both intensity and phase fluctuations.
We emphasize that the mechanism behind (\ref{eq:Vna}) is related
to nonadiabatic corrections, as described by Olshanii and Dum \cite{Olshanii1996},
and is conceptually different from schemes relying on a substructuring
AC Stark based OLs by radio frequency or laser fields \cite{yi2008state,nascimbene2015dynamic},
or in combination with DSs \cite{Zubairy2008}. Fig.~2a is a plot
of the BO potentials $E_{0,\pm}(x)$ for blue detuning $\Omega=\Delta>0$
and $\epsilon=0.16$ with parameters chosen to illustrate the main
features (with similar results for red detuning). For $\Delta\gg\Omega_{c,p}$,
the bright state (BS) $\ket{E_{+}(x)}\rightarrow\sin\alpha(x)\ket{g_{1}}+\cos\alpha(x)\ket{g_{2}}$
corresponds to the standard OL $E_{+}(x)\rightarrow\frac{\hbar}{4}[\Omega_{p}^{2}+\Omega_{c}^{2}(x)]/(\Delta+\frac{i}{2}\Gamma)$
as a second order Stark shift (c.f.~Fig.~2a).

\begin{figure}
\includegraphics[clip,width=8.6cm]{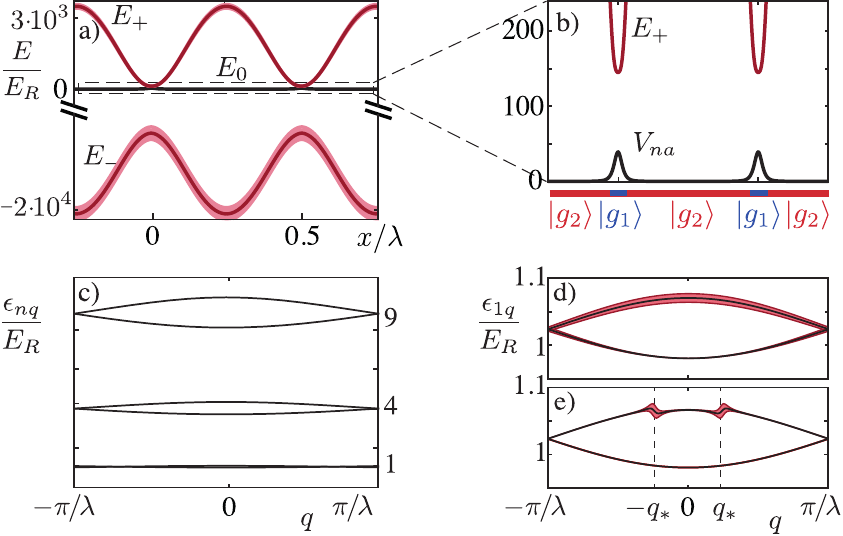} \caption{(Color online) a) BO potentials $E_{0}(x)=0$ (black line) and $E_{\pm}(x)$
(red line, with line width indicating $\textrm{Im}\,\,E_{\pm}(x)$)
for DS and BS, respectively. Parameters: $\Omega_{c}=\Delta=1.7\times10^{4}E_{R}$,
$\epsilon=0.16$, $\Gamma=5\times10^{2}E_{R}$. b) Zoom showing $V_{na}(x)$
(black) and the lower part of $E_{+}(x)$ (red). c) Lowest Bloch bands
$\epsilon_{n,q}$ for $V_{na}(x)$ for Brillouin zone $\left|q\right|<\pi/\lambda$
(see text); $\epsilon=0.1$. d), e) Zooms of the lowest Bloch band
of the DS lattice, including couplings to BSs for $\epsilon=0.1,\Omega_{p}=2\times10^{3}E_{R}$,
$\Gamma=10^{3}E_{R}$ (d), and $\Gamma=10E_{R}$ (e) (see text and
\cite{SM}). Black lines indicate $\textrm{Re}\,\,\epsilon_{1,q}$,
and red shadings the widths $\textrm{Im}\,\,\epsilon_{1,q}=-\hbar\gamma_{1,q}/2$.}
\end{figure}

To quantify the above discussion and assess the validity of the BO
approximation we present now a derivation and analysis of optical
potentials arising from nonadiabatic corrections to atomic motion,
and effects of spontaneous emission (due to admixture of bright channels).
Expanding the atomic wavefunction in the BO channels, $\ket{\Psi(x)}=\sum_{\sigma}\Psi_{\sigma}(x)\ket{E_{\sigma}(x)}$,
results in the coupled channel equation for $\Psi_{\sigma}(x)$ \cite{Olshanii1996,Ruseckas2005}.
The corresponding Hamiltonian is ${\cal H}=(-i\hbar\partial_{x}-A(x))^{2}/2m+V(x)$,
where the diagonal matrix $V_{\mu\sigma}(x)=E_{\sigma}(x)\delta_{\mu\sigma}$
contains BO potentials (see Fig.~2). Nonadiabatic processes, coupling
the BO channels, arise from the spatial variation of the internal
eigenstates, $-i\hbar\partial_{x}\ket{E_{\sigma}(x)}=\sum_{\mu}\ket{E_{\mu}(x)}A_{\mu\sigma}(x)$
with scaling $A_{\mu,\sigma}\sim\hbar/\ell$ (see \cite{SM}).

We are interested in the regime of approximate adiabatic decoupling
of BO channels. This requires that the separations between DS and
BS are larger than the channel couplings. For the DS, the lowest order
contribution from the $A^{2}$-term gives rise to the nonadiabatic
(conservative) potential (\ref{eq:Vna}) (see Fig.~2b), with consistency
requirement $V_{na}(x)\ll{\rm min}|E_{\pm}(x)|$. We discuss this
by setting the external potential $V(x)=V_{L}(x)=0$, and studying
the 1D bandstructure for the $\Lambda$-scheme of Fig.~1b. We compare
below the results for (i) the single-channel DS potential $V_{na}(x)$
with (ii) the exact diagonalization of the (non-Hermitian) Hamiltonian
$H$, using a Bloch ansatz $\Psi(x)=e^{iqx}u_{n,q}(x)$ with quasimomentum
$q$ to obtain the (complex) energies $\epsilon_{n,q}$. In the first
case we have a unit cell $\lambda/2$ and thus a Brillouin zone $|q|<2\pi/\lambda$,
while $H$ has periodicity $\lambda$, and thus $|q|<\pi/\lambda$,
so that the bands of the first case appear as `folded back' in the
second case (see Figs. 2c,d,e).

For the DS potential $V_{na}(x)$ the band structure is for $\epsilon\ll1$
analogous to that of a Kronig-Penney model \cite{Merzbacher1998}.
For the lowest Bloch bands $n=1,2\ldots$ in the DS channel $0$ we
obtain (see \cite{SM}) 
\begin{equation}
\epsilon_{n,q}^{(0)}\!\approx\!E_{R}n^{2}\!\left\{ 1\,+\!\frac{4\epsilon}{\pi^{2}}\left[1\!+\!(-1)^{n}\!\cos\frac{\pi q}{k}\right]\right\} ,\,|q|\le\frac{2\pi}{\lambda}\label{eq:eps}
\end{equation}
in very good quantitative agreement with the band structure obtained
from $H$. These bands have narrow width $\sim\epsilon$, corresponding
to a hopping amplitude $J_{n}=2E_{R}\epsilon n^{2}/\pi^{2}$ in the
terminology of the tight-binding Hubbard model \cite{bloch2008many}.
The energy offset of these bands $E_{R}n^{2}$ is close to the levels
in the infinitely deep rectangular well of the width $\lambda/2$,
with anharmonic band spacing (independent of $\epsilon$, and thus
the height of the potential). The Wannier functions associated with
these bands resemble the eigenfunctions of a box potential. This is
in marked contrast to the band structure in a $V(x)=V_{0}\sin^{2}(kx)$
OL, where energies of low lying bands are harmonic oscillator-like,
and the Wannier functions are strongly localized $a_{L}\ll\lambda/2$
(Lamb-Dicke regime) \cite{bloch2008many}. The spectroscopy of these
bands could be investigated with time-of-flight, and by modulating
the lattice. A discussion of this and of loading the lowest Bloch
band can be found in \cite{SM}.

For the DS channel $0$, the nonadiabatic couplings to the bright
(dissipative) BO channels $\pm$ result in a small correction $\delta\epsilon_{n,q}$
to the dispersion, which contains a imaginary part $\text{Im}\,\delta\epsilon_{n,q}=-\hbar\gamma_{n,q}/2<0$
signalling decay of atoms in the Bloch band. Figs. 2d,e indicate the
width of the lowest Bloch band $\gamma_{1,q}$, by a red shaded region
around the dispersion relation $\epsilon_{1,q}$, as obtained from
a numerical solution of the coupled channel equations. The parameters
are chosen to illustrate the limits of a `large' and `small' decay
width $\Gamma$ (see \cite{SM}). We note that in both cases $\gamma_{1,q}\ll J_{1}$,
i.e.~dissipative corrections are typically very small, while $\gamma_{1,q}$
shows a nontrivial $q$-dependence. In Fig.~2d we can parametrize
\begin{equation}
\gamma_{1,q}\approx\gamma_{1}\sin^{2}(\pi q/2k),\quad|q|\leq\frac{2\pi}{\lambda}\label{eq:g0}
\end{equation}
(see \cite{SM}), where for the lowest Bloch band the decay increases
with $q$ (Fig.~2d) - something we expect from a %
\mbox{%
STIRAP%
} scenario \cite{vitanov2016stimulated}, where faster atomic motion
leads to a stronger violation of adiabaticity and thus depopulation
of the DS. In contrast, Fig.~2e shows the appearance of \emph{resonances}
in $q$: as discussed in \cite{SM} these appear when for a given
$q=q_{\star}$ the energies in the dark channel $0$ becomes energetically
degenerate with energies in the bright open channel $-$, $\epsilon_{n,q_{\star}}^{(0)}\approx\textrm{Re}\,\epsilon_{n,q_{\star}+k}^{(-)}$.
With increasing $\Gamma$ relative to the strength of the nonadiabatic
couplings these resonance get successively washed out, transitioning
to the generic behavior of Fig.~2d. We refer to \cite{SM} for a
detailed discussion of $\gamma_{n,q}$, and in particular scaling
with system parameters.

Returning to Fig.~1a we point out that the above discussion can be
generalized to DSs in 2D and 3D configurations. Thus we can replace
$\Omega_{c}(x)\rightarrow\Omega_{c}(x,y)$ and $V_{L}(x)\rightarrow V_{L}(x,y)$,
while preserving the existence of a DS $\ket{E_{0}(x,y)}$, allowing
to add a (standard) OL for motion in the $y$-direction, or the realization
of an atomic double wire with separation $\ell$.

\begin{figure}[!]
\centering \includegraphics[width=8.6cm]{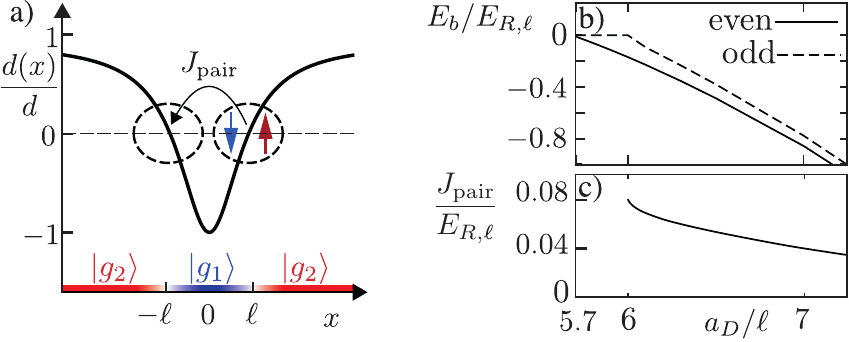} \caption{(Color online) a) Position-dependent dipole moment $d(x)/d$, for
$-d_{1}=d_{2}=d$. Molecules exist as bound states of two atoms due
to dipolar attraction at the interfaces, where $d(x)$ changes sign.
b) Bound state energies $E_{b}$, and c) hopping amplitude $J_{{\rm pair}}$
of molecules between the two interfaces for $\epsilon=0.1$ in units
of $E_{R,\ell}\equiv\hbar^{2}/2m\ell^{2}=E_{R}/\epsilon^{2}$. }
\label{fig:Fig3} 
\end{figure}

We now turn to a study of quantum many-body physics, and discuss as
an illustrative example motion of two atoms confined in the subwavelength
structure of Fig.~1, and interacting via magnetic dipolar interactions.
The validity of (\ref{eq:Vna}) in a many-body Schrödinger equation
will be discussed below. We assume that the two-body physics can be
modeled by the external motion of each atom governed by $V_{L}(x)+V_{na}(x)$,
while the internal state is the BO channel $\ket{E_{0}(x)}$, with
the unique feature of an $x$-dependent internal state (see bottom
of Fig.~1a). We consider two ground states (spins) with associated
magnetic dipole moments $d_{1},d_{2}$, oriented according to a quantization
axis defined by an external magnetic field, so that each atom acquires
an effective position-dependent dipole moment, $d(x)=d_{1}\cos^{2}\alpha(x)+d_{2}\sin^{2}\alpha(x)$.
Fig.~3a is a plot of this dipole moment for a choice of states with
$-d_{1}=d_{2}\equiv d$, with the spatial variation of $\ket{E_{0}(x)}$
now imprinted as a variation of $d(x)$ on the scale $\ell$. The
magnetic (dipolar) interactions between the atoms is thus modulated
by this spatial dependence. There are two generic situations, the
first (i) with the dipole moments oriented along $x$, and the second
(ii) with dipole oriented perpendicular. In the first case, two atoms
on opposite sides of the barrier in the double layer attract each
other in a head-to-tail configuration. For the case of electric dipole
moments as realized with polar molecules, stored in a 2D double layer
from a OL with $\lambda/2$ separation, the formation of bound states
as building block for quantum phases has been studied \cite{Wang:2006ja,Santos2010,Zoller2011,Baranov:ChemRev}.
Here we note that this physics of strong interactions becomes accessible,
when the dipolar length, $a_{D}=md^{2}/\hbar^{2}$ \cite{Ferlaino2016,Lev2016}
characterizing the dipolar interactions \cite{footd}, is comparable
to the average distance between the atoms (here $\sim a_{L}$ with
$\ell\ll a_{L}\ll\lambda_{L}$).

Instead, we focus here on physics of perpendicular dipole moments
(ii) at the \emph{interface} between the spin structure, $\ket{g_{1}}\leftrightarrow\ket{g_{2}}$,
as shown in Fig.~3. If the dipoles are oriented perpendicular to
$x$, atoms on opposite sites of the interface attract each other,
thus allowing for the formation of a bound state as a `domain wall'
molecule. The situation is illustrated by the following two-particle
Hamiltonian (see \cite{SM} for detailed description):
\begin{equation}
H\!=\!\sum_{i=1,2}\!\left[-\frac{\hbar^{2}\partial_{x_{i}}^{2}}{2m}\!+\!V_{na}(x_{i})\right]\!+\!\frac{d(x_{1})d(x_{2})}{|x_{1}\!-\!x_{2}|^{3}}\label{eq:H2}
\end{equation}
with $d(x)$ modulated on the scale $\ell$, assuming strong confinement
$\ell_{\perp}<\ell$ in the transverse plane (and setting $V_{L}=0$).
According to Fig.~3 we find that the requirement for a bound state
of size $\ell$ to form is $a_{D}/\ell\sim6$ \cite{Ye2013,Jin2015,footb},
where the `domain wall' molecules sit on the slope of the nonadiabatic
potential. These molecules exist at both the left and right interfaces
$\pm\ell$, and can hop between them, realizing a double layer with
subwavelength distance. The (potentially large) amplitude $J_{{\rm pair}}$
for hopping is reflected as a hybridization of molecular orbitals
on the left and right interfaces, seen in Fig.~3 as a splitting between
the even and odd states. We can also obtain \emph{trimers} as bound
states of three atoms, where two spin-up dipoles sit to the left (right)
of $-\ell$ ($+\ell$) and a spin-down in the middle.

From an atomic physics point of view, a $\Lambda$-scheme and a nonadiabatic
DS potential can be realized with both Alkali and Alkaline Earth atoms,
where two ground states are chosen from a Zeeman or hyperfine manifold.
Remarkably, these nonadiabatic potentials exist, on the level of single-atom
physics, as \emph{conservative} optical potentials even \emph{on-resonance}
$(\Delta=0)$ and for short lived excited states (but still $\Omega_{c,p}\gg\Gamma$).
In going off-resonance the nonadiabatic conservative potential will
persist albeit with an increasing requirement for laser power to satisfy
the adiabaticity requirement, in particular $V_{na}(0)<\frac{\hbar}{4}\Omega_{p}^{2}/\Delta$
($\Omega_{p}\ll\Delta)$ for $\Delta>0$ as shown in Figs.~2a,b.
With increasing detuning the three-level model will eventually break
down, and the coupling to several excited states may become important.
This situation parallels the challenges in realizing spin-dependent
OLs \cite{Mandel:2003ej,Lee:2007ip,SoltanPanahi:2011ey,Dai:us}, and
spin-orbit coupling in $\Lambda$-systems with Alkali atoms \cite{Lin:2011hn,Wang:2012gv,Cheuk:2012tl,Stuhl:2015cb,Huang:2016kf,Spielman2013,Dalibard:2011gg,Goldman:2016tw},
where the electronic spin-flip implicit in coupling two ground states
via Raman transition is suppressed for detunings larger than the fine
structure splitting of the excited state. We note, however, the encouraging
prospects provided by Lanthanides in realizing spin-orbit couplings
\cite{Lev2016,Cui:2013ki,Mancini:2015fb} and synthetic gauge fields
\cite{Nascimbene:2013to,Wall:2016cl}. In a many-atom context, going
to off-resonant laser excitation is a necessary requirement to suppress
inelastic collision channels (involving laser excitation at the Condon
point), and we expect a similar requirement here. As discussed in
the context of polar molecules, long range repulsive dipolar interactions
in combination with low-dimensional trapping (1D or 2D) can provide
a shield in atom-atom collisions at low energies \cite{footc}, thus
suppressing inelastic loss and instabilities from short range physics
\cite{Micheli:2010jq}. %This motivates the study of model system illustrating the physics of two atoms interacting via magnetic dipole-dipole interactions in the subwavelength structure of Fig.~1 (see Fig.~3). 
 
\begin{figure}
\includegraphics[clip,width=8.6cm]{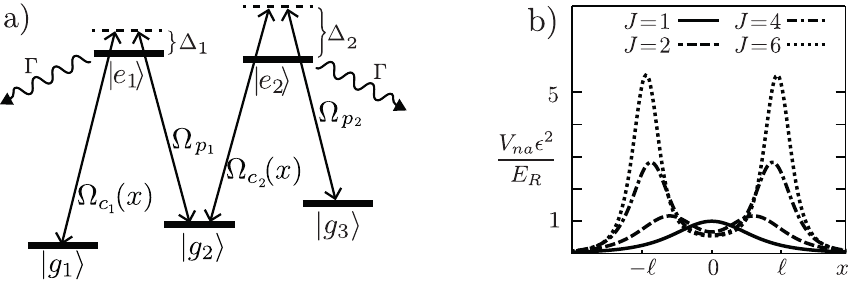} \caption{ a) Atomic zig-zag (double-$\Lambda$) configuration with $\Omega_{c_{i}}(x)$
strong standing waves and $\Omega_{p_{i}}$ weak probe beams, and
b) the corresponding {\em nonadiabatic optical potentials} on the
subwavelength scale $\ell\ll\lambda$ for an atomic angular momentum
$J_{g}=J_{e}\equiv J$ transition, where the Zeeman levels are coupled
by circularly polarized laser fields. With increasing $J$ a double
barrier structure develops. }
\end{figure}

To conclude, $\Lambda$-type configurations with nonadiabatic DS optical
potentials \cite{note} are readily generalized to zig-zag configurations
as in Fig.~4a (see also \cite{SM}). This yields a \emph{double-peaked
structure} on the scale $\ell$ as in Fig.~4b. These ideas enable
writing complex spatial spin patterns \cite{Zubairy2008} and associated
landscapes of nonadiabatic potentials. On the many-atom level spatially
varying internal structures result in \emph{position-dependent} interparticle
interactions. This provides a novel setting for many-body atomic systems,
illustrated here for magnetic dipole-dipole interactions, and poses
interesting questions as quantum chemistry in atomic collisions at
subwavelength distances. 
\begin{acknowledgments}
We thank J. Budich, L. Chomaz, J. Dalibard, M. Dalmonte, F. Ferlaino,
R. Grimm, T. Pfau, and T. Porto for helpful comments. Work at Innsbruck
is supported by the ERC Synergy Grant UQUAM, the Austrian Science
Fund through SFB FOQUS, and EU FET Proactive Initiative SIQS. H.~P.
was supported by the NSF through a grant for the Institute for Theoretical
Atomic, Molecular, and Optical Physics at Harvard University and the
Smithsonian Astrophysical Observatory. 
\end{acknowledgments}

\appendix
%dummy comment inserted by tex2lyx to ensure that this paragraph is not empty%dummy comment inserted by tex2lyx to ensure that this paragraph is not empty%dummy comment inserted by tex2lyx to ensure that this paragraph is not empty%dummy comment inserted by tex2lyx to ensure that this paragraph is not empty%dummy comment inserted by tex2lyx to ensure that this paragraph is not empty
 \bibliographystyle{apsrev4-1.bst}

\clearpage

\appendix
\section*{Supplemental Material}

\section{Born-Oppenheimer approach}

\label{app:BO}

Here we present details of the Born-Oppenheimer (BO) approach used
for the analysis of the $\Lambda$-system described in the main text.
In the bare atomic basis $\left\vert g_{1}\right\rangle $, $\left\vert e\right\rangle $,
and $\left\vert g_{2}\right\rangle $, the Hamiltonian of the system
reads 
\begin{equation}
H=-\frac{\hbar^{2}\partial_{x}^{2}}{2m}+\hbar\left(\begin{array}{ccc}
0 & \Omega_{c}(x)/2 & 0\\
\Omega_{c}(x)/2 & -\widetilde{\Delta} & \Omega_{p}/2\\
0 & \Omega_{p}/2 & 0
\end{array}\right)\label{H1}
\end{equation}
with $\tilde{\Delta}=\Delta+i\Gamma/2$ This Hamiltonian is non-Hermitian
(for $\Gamma\neq0$) and has complex eigenvalues and a bi-orthogonal
set of the left and right eigenstates. In the BO approximation we
drop the kinetic energy term, and corresponding right (adiabatic)
eigenstates are the dark $\left\vert E_{0}\right\rangle $ and bright
$\left\vert E_{\pm}\right\rangle $ eigenstates 
\begin{align*}
\left\vert E_{0}\right\rangle  & =-\cos\alpha\left\vert g_{1}\right\rangle +\sin\alpha\left\vert g_{2}\right\rangle ,\\
\left\vert E_{\pm}\right\rangle  & =N_{\pm}\left\{ \epsilon_{\pm}\left\vert e\right\rangle +\left[\sin\alpha\left\vert g_{1}\right\rangle +\cos\alpha\left\vert g_{2}\right\rangle \right]\right\} 
\end{align*}
with $\alpha(x)=\arctan[\Omega_{c}(x)/\Omega_{p}]$ have eigenenergies
$E_{0}=0$ and $E_{\pm}(x)$, respectively. Here $\epsilon_{\pm}=2E_{\pm}(x)/E(x)$
with $E(x)=\hbar\sqrt{\Omega_{p}^{2}+\Omega_{c}(x)^{2}}$, and $N_{\pm}=(1+\epsilon_{\pm}^{2})^{-1/2}$.
We emphasize that these eigenstates and eigenenergies depend parametrically
on the position $x$.

Straightforward calculations of the derivatives of the eigenstates
$\left|E_{\sigma}(x)\right\rangle $ for $\sigma=0,\pm$ in the expression
$-i\hbar\partial_{x}\left\vert E_{\sigma}\right\rangle =\sum_{\mu}\left\vert E_{\mu}\right\rangle A_{\mu\sigma}$
give 
\[
A=-i\hbar\alpha^{\prime}\left(\begin{array}{ccc}
0 & -N_{+} & -N_{-}\\
N_{+} & 0 & -C\\
N_{-} & C & 0
\end{array}\right)
\]
with 
\[
C=\widetilde{\Delta}\frac{\Omega_{c}(x)}{2\Omega_{p}}\frac{E(x)}{E(x)^{2}+\widetilde{\Delta}^{2}},
\]
such that the Hamiltonian $H$ for the wave functions $\Psi_{\sigma}(x)$
reads 
\begin{align}
{\cal H} & =-\frac{\hbar^{2}\partial_{x}^{2}}{2m}+\left(\begin{array}{ccc}
0 & 0 & 0\\
0 & E_{+} & 0\\
0 & 0 & E_{-}
\end{array}\right)\label{eq:HD1}\\
& +\frac{\hbar^{2}(\alpha^{\prime})^{2}}{2m}\left(\begin{array}{ccc}
1 & 0 & 0\\
0 & N_{+}^{2}+C^{2} & 0\\
0 & 0 & N_{-}^{2}+C^{2}
\end{array}\right)\label{eq:HD2}\\
& -\frac{i\hbar}{2m}\left(\partial_{x}A+A\partial_{x}\right)\label{eq:HC1}\\
& -\frac{\hbar^{2}(\alpha^{\prime})^{2}}{2m}\left(\begin{array}{ccc}
0 & CN_{-} & -CN_{+}\\
CN_{-} & 0 & -N_{+}N_{-}\\
-CN_{+} & -N_{+}N_{-} & 0
\end{array}\right).\label{eq:HC2}
\end{align}
The Hamiltonian is naturally split into a diagonal part ${\cal H}_{D}$
(the first two lines) which determines the BO bandstructures for the
dark and bright channels, and off-diagonal terms ${\cal H}_{C}$ (the
last two lines) determining the coupling between them. The coupling
contains $\alpha^{\prime}(x)$ and, therefore, mostly takes place
in the narrow regions with the width $\Delta x\sim\epsilon/k\sim\epsilon\lambda$
around the points $x_{s}=\pi s/k=s\lambda/2$ with integer $s$. The
strength of the coupling is essentially determined by the ratio $\kappa$
of the strength of the non-adiabatic potential $V_{na}(x=0)=E_{R}/\epsilon^{2}$
and the energy gap between the dark and the bright states. The latter
depends on the detuning and is of the order of $\Omega_{p}$ for the
resonant case $\Delta=0$, and $\Omega_{p}^{2}/\left\vert \Delta\right\vert $
for the off-resonant case with large $\Delta$. As a result, for the
considered case $\kappa\ll1$, where $\kappa=E_{R}/\Omega_{p}\epsilon^{2}$
for $\Delta=0$ and $\kappa=E_{R}\left\vert \Delta\right\vert /\Omega_{p}^{2}\epsilon^{2}$
for large $\left\vert \Delta\right\vert $, the couplings are small
justifying the perturbative approach in the main text.

\section{Band structure in the Born-Oppenheimer approximation}

\label{app:band}

In the following we provide details on the band structure of the dark
and bright channels in the BO approximation. We first analyze the
case of uncoupled BO channels and then discuss the effects of non-adiabatic
couplings between them.

The diagonal Hamiltonian describing uncoupled BO dark and bright channels
reads 
\[
{\cal H}_{D}=-\frac{\hbar^{2}\partial_{x}^{2}}{2m}+\left(\begin{array}{ccc}
V_{na}(x) & 0 & 0\\
0 & E_{+}(x) & 0\\
0 & 0 & E_{-}(x)
\end{array}\right),
\]
where for $\kappa\ll1$ we neglect the nonadiabatic contribution to
the bright channels.

The second term in ${\cal H}_{D}$ is periodic with $\lambda/2=\pi/k$,
and, therefore, the single-particle eigenstates are Bloch wave functions
$\psi_{n,q}(x)$ characterized by the quasimomentum $q$ in the Brillouin
zone, $q\in[-k,k]$, such that $\psi_{n,q}(x+\lambda/2)=\exp(iq\lambda/2)\psi_{n,q}(x)$,
and by the band index $n=1,2,\ldots$, with the corresponding eigenenergies
$\epsilon_{n,q}$ forming the band structure. Note that the periodicity
of the Hamiltonian ${\cal H}$ is $2\pi/k=\lambda$ (because of the
coupling terms proportional to $\alpha^{\prime}$), while the diagonal
part ${\cal H}_{D}$ has periodicity $\pi/k=\lambda/2$ ($\alpha^{\prime}$
changing sign under $x\rightarrow x+\pi/k$). For this reason, for
describing the band structure we will use both the Brillouin zone
(BZ) of ${\cal H}_{D}$ with quasimomenta $q\in[-k,k]$ and the (folded)
Brillouin zone (FBZ) of ${\cal H}$ with $\overline{q}\in[-k/2,k/2]$,
with the obvious mapping between them: $\overline{q}=q$ for $q\in[-k/2,k/2]$
and $\overline{q}=q-k$ or $\overline{q}=q+k$ for $q\in[k/2,k]$
or $q\in[-k,-k/2]$, respectively. Under this folding, the dispersion
relation $\epsilon_{n,q}$ for the Bloch states in BZ is mapped on
the \emph{continuous} dispersion relation $\overline{\epsilon}_{n,\overline{q}}$
for states in FBZ, which has two branches $\epsilon_{n,\overline{q}}$
and $\epsilon_{n,\overline{q}+k}$ continuously matching at $\overline{q}=\pm k/2$.

\subsection{Band structure of the dark-state channel $0$}

The part of the Hamiltonian ${\cal H}_{D}$ for the dark-state 
\[
{\cal H}_{0}=-\frac{\hbar^{2}\partial_{x}^{2}}{2m}+V_{na}(x),
\]
is Hermitian and describes the motion of particle in the periodic
(with $\pi/k$) set of sharp $\delta$-like potential peaks of the
width $\Delta x\sim\epsilon/k\ll\pi/k$ around points $x_{n}=(\pi/k)n$,
and height $E_{R}/\epsilon^{2}\gg E_{R}$ (analog of the Kronig-Penney
model). In this case, the structure of the low-energy bands with $\epsilon_{n,q}\ll\hbar^{2}/m(\Delta x)^{2}$
is fully determined by the transmission amplitude $t(E)$ through
a single barrier \cite{Merzbacher1998}. The corresponding relation
between the quasimomentum $q$ and the energy $\epsilon_{n,q}$ reads
\begin{equation}
\cos\pi q/k=\frac{1}{2}\left[\frac{1}{t^{\ast}(\epsilon_{n,q})}e^{i\pi Q/k}+\frac{1}{t(\epsilon_{n,q})}e^{-i\pi Q/k}\right],\label{eigenvalue equation}
\end{equation}
where $Q=\sqrt{2m\epsilon_{n,q}/\hbar^{2}}$ and we took into account
that the length of the unit cell is $\pi/k$.

To find the transition amplitude $t(E)$ as a function of energy $E$
we have to solve the scattering problem for a single barrier located
at $x=0$, which for $\epsilon\ll1$ can be approximated as 
\begin{equation}
V_{na}(x)\approx V(x)=E_{R}\frac{\epsilon^{2}}{[\epsilon^{2}+(kx)^{2}]^{2}}.\label{eq:Vna_approx}
\end{equation}
The corresponding Schrödinger equation reads 
\[
\left[-\frac{\hbar^{2}\partial_{x}^{2}}{2m}+V(x)\right]\psi(x)=E\psi(x),
\]
and the boundary conditions are 
\begin{align*}
\psi(x & \rightarrow-\infty)=e^{iQx}+r(E)e^{-iQx},\\
\psi(x & \rightarrow+\infty)=t(E)e^{iQx},
\end{align*}
where $Q=\sqrt{2mE/\hbar^{2}}$. In the units $s=kx/\epsilon$ we
get 
\[
\left[-\frac{d^{2}}{ds^{2}}+\frac{1}{(1+s^{2})^{2}}\right]\psi(s)=\epsilon^{2}\left(\frac{Q}{k}\right)^{2}\psi(s).
\]
Inside the barrier ($\left\vert s\right\vert \ll1/\sqrt{\epsilon}$
or $\left\vert x\right\vert \ll\sqrt{\epsilon}/k$) we can neglect
the right-hand-side and reduce the equation to the form 
\[
\left[-\frac{d^{2}}{ds^{2}}+\frac{1}{(1+s^{2})^{2}}\right]\psi(s)=0
\]
which general solution is 
\begin{equation}
\psi(s)=\sqrt{1+s^{2}}(A+B\arctan s),\quad\left\vert s\right\vert \ll\frac{1}{\sqrt{\epsilon}}.\label{solution in}
\end{equation}
Outside the barrier ($\left\vert s\right\vert \gg1/\sqrt{\epsilon}$),
the potential is negligible, and the resulting equation 
\[
-\frac{d^{2}}{ds^{2}}\psi(s)=\epsilon^{2}\left(\frac{Q}{k}\right)^{2}\psi(s)
\]
has general solution 
\begin{equation}
\psi(s)=C\exp\left(\frac{i\epsilon Qs}{k}\right)+D\exp\left(-\frac{i\epsilon Qs}{k}\right),\quad\left\vert s\right\vert \gg\frac{1}{\sqrt{\epsilon}}.\label{solution out}
\end{equation}
For the scattering problem we have to set $C=1$, $D=r(E)$ for large
negative $s$, and $C=t(E)$, $D=0$ for large positive $s$, respectively.
After matching Eqs. (\ref{solution in}) and (\ref{solution out})
at $\left\vert s\right\vert =\left\vert s_{\ast}\right\vert \sim1/\sqrt{\epsilon}$
($1\ll\left\vert s_{\ast}\right\vert \ll1/\epsilon$) where the barrier
potential is comparable with the energy, we obtain 
\begin{equation}
t(E)=-\left(1-i\frac{\pi}{2\epsilon}\frac{k}{Q}\right)^{-1}\label{transition amlitude}
\end{equation}

This result, although being similar, is not identical with that for
the $\delta$-functional potential $V_{\delta}(x)=(\hbar^{2}/2m)(\pi k/2\epsilon)\delta(x)$
of the strength $\int dxV(x)$. The latter is equivalent to imposing
the boundary conditions at the origin

\[
\begin{aligned}\psi_{+}-\psi_{-} & =0,\\
\psi'_{+}-\psi'_{-} & =\frac{\pi k}{4\epsilon}(\psi_{+}+\psi_{-}),
\end{aligned}
\]
where $\psi_{\pm}=\psi(\pm0)$ and $\psi'_{\pm}=\partial_{x}\psi(\pm0)$
are the values of the wave function and its derivative, respectively,
on the left ($x=-0$) and on the right ($x=+0$) of the $\delta$-functional
barrier $V_{\delta}(x)$. Similar boundary conditions can also be
written for the potential $V(x)$, although they are applicable only
for the wave functions with energies $E\ll E_{R}/\epsilon^{2}$: After
writing the asymptotic of the wave function in the form $\psi(x)\to\psi_{\pm}+\psi'_{\pm}x$
for $x\sim\pm\sqrt{\epsilon}/k$ and matching them with the asymptotic
of Eqs. (\ref{solution in}), we obtain 
\[
\begin{aligned}\psi_{+}+\psi_{-} & =0,\\
\psi'_{+}+\psi'_{-} & =-\frac{\pi k}{2\epsilon}(\psi_{+}-\psi_{-}).
\end{aligned}
\]
To demonstrate the difference between the two sets of boundary conditions,
we present the result for the transition amplitude $t_{\delta}(E)$
through the potential $V_{\delta}(x)$: 
\begin{equation}
t_{\delta}(E)=\left(1+i\frac{\pi}{4\epsilon}\frac{k}{Q}\right)^{-1},\label{eq:transition delta}
\end{equation}
which has similar scaling for $Q\to0$ as $t(E)$ but different coefficient
and subleading terms.

\subsection{Dispersion relation for the Bloch bands}

With the expression (\ref{transition amlitude}) for the transition
amplitude, Eq.~(\ref{eigenvalue equation}) for the dispersion relation
for the Bloch bands reads 
\[
\cos\pi q/k=-\cos(\pi Q/k)+\frac{\pi k}{2Q\epsilon}\sin(\pi Q/k).
\]
For $\epsilon\ll1$, the solutions for $Q$ are located near the points
$kn$, $n=1,2,\ldots$ After linearizing around these points we obtain
for $1\leq n<\epsilon^{-1}$ 
\[
Q_{n}(q)\approx kn\left\{ 1+\frac{2\epsilon}{\pi^{2}}\left[1+(-1)^{n}\cos\frac{\pi q}{k}\right]\right\} 
\]
and $\hbar^{2}Q_{n}^{2}(q)/2m$ gives Eq.~(2) in the main text for
the dispersion of the lower Bloch bands in the dark-state channel.
Note that if we approximate the non-adiabatic potential with the periodic
set of $\delta$-functions {[}in other words, use $t_{\delta}(E)$,
Eq.~(\ref{eq:transition delta}), instead of $t(E)$ in Eq. (\ref{transition amlitude}){]},
the expression (2) in the main text for the dispersion relation changes
into 
\begin{equation}
\epsilon_{n,q}^{(\delta)}\approx E_{R}n^{2}\left\{ 1-\frac{8\epsilon}{\pi^{2}}\left[1-(-1)^{n}\cos\frac{\pi q}{k}\right]\right\} .\label{eq:epsilon for delta}
\end{equation}

\subsubsection{Wave functions for the Bloch band}

Eqs. (\ref{solution in}) and (\ref{solution out}) can also be used
for finding the Bloch wave functions $\psi_{n,q}(x)$, as we demonstrate
for the lowest band $n=1$ with energies $\epsilon_{1,q}\simeq E_{R}$.
The wave functions for higher bands can be found in the same way,
and the answer will be given at the end of this section). The expression
for $\psi_{1,q}(x)$ will be given for $q\in[-k,k]$ and $x\in[-\pi/2k,\pi/2k]$.
For other values of $x$, the wave function can be calculated from
the relation $\psi_{1,q}(x+\pi/k)=\exp(i\pi q/k)\psi_{1,q}(x)$.

With the approximation $V_{na}(x)\approx V(x)$, Eq. (\ref{eq:Vna_approx}),
and new variable $s=kx/\epsilon\in[-\pi/2\epsilon,\pi/2\epsilon]$,
the equation for $\psi_{1,q}(s)$ reads 
\[
\left[-\frac{d^{2}}{ds^{2}}+\frac{1}{(1+s^{2})^{2}}\right]\psi_{q}(s)=\epsilon^{2}\psi_{1,q}(s),
\]
where we used the fact that $\epsilon_{1,q}\simeq E_{R}$. The two
solutions (\ref{solution in}) and (\ref{solution out}) have to be
matched at $\left\vert s_{\ast}\right\vert \sim1/\sqrt{\epsilon}$
in such a way that 
\begin{equation}
\frac{\psi_{1,q}(\pi/2\epsilon)}{\psi_{1,q}(-\pi/2\epsilon)}=\exp(i\pi q/k).
\end{equation}
This gives in the original variable $x\in[-\pi/2k,\pi/2k]$ 
\begin{align}
\psi_{1,q}(x) & \approx N\sqrt{\epsilon^{2}+\sin^{2}(kx)}\left[\cos\left(\frac{\pi q}{2k}\right)\right.\nonumber \\
& \left.+\frac{2i}{\pi}\sin\left(\frac{\pi q}{2k}\right)\arctan\left(\frac{\tan(kx)}{\epsilon}\right)\right]\label{eq:wave function dark}\\
& =\cos\left(\frac{\pi q}{2k}\right)\psi_{0}(x)+i\sin\left(\frac{\pi q}{2k}\right)\psi_{\pi}(x)\label{wave function parts}
\end{align}
for $q\in[-k,k]$, where $N$ is the normalization coefficient, $N\approx\sqrt{2k/\pi}$,
corresponding to the unity of the integral of $\left\vert \psi_{1,q}(s)\right\vert ^{2}$
over the unit cell $x\in[-\pi/2k,\pi/2k]$. Note the substitution
$\tan(kx)/\epsilon$ instead of $s=kx/\epsilon$ in the argument of
the $\arctan$ function, which provides the second independent solution
{[}$\sim\cos(kx)${]} outside the barrier in addition to $\sin(kx)$.

For the wave functions of the higher bands similar considerations
give the following approximate expression: 
\begin{equation}
\begin{aligned}\psi_{n,q}(x) & \approx N_{n}\sqrt{\epsilon^{2}+\sin^{2}z}\left[-\frac{\sin(nz)}{\sin z}\left(\frac{z}{i|z|}\right)^{n+1}e^{i\frac{\pi q}{2k}\frac{z}{|z|}}\right.\\
& -\left.\frac{2in}{\pi}\cos\left(\frac{\pi q}{2k}-\frac{\pi n}{2}\right)\cos(nz)\arctan\left(\frac{\epsilon}{\tan z}\right)\right],
\end{aligned}
\label{eq:Bloch wf n}
\end{equation}
where $z\equiv kx\in[-\pi/2,\pi/2]$, $q\in[-k,k]$, and the band
index $n$ should satisfy the condition $n\epsilon\ll1$.

\subsubsection{Band structure of the bright-state channels $\pm$}

The Hamiltonians for the bright-state BO channels 
\[
{\cal H}_{\pm}=-\frac{\hbar^{2}\partial_{x}^{2}}{2m}+E_{\pm}(x)
\]
describe the motion in the complex potentials of local bright-state
eigenenergies.

For the near-resonant case $\Delta=0$, we have 
\[
E_{\pm}(x)\approx\pm\frac{1}{2}E(x)-\frac{i\Gamma}{4},
\]
while for the off-resonant case $\left\vert \Delta\right\vert \gg\Omega_{c}$
\begin{align*}
E_{+}(x) & \approx\frac{E^{2}(x)}{4\widetilde{\Delta}}\approx\frac{E^{2}(x)}{4\Delta}\left(1-i\frac{\Gamma}{2\Delta}\right),\\
E_{-}(x) & \approx-\Delta-i\frac{\Gamma}{2}-\frac{E^{2}(x)}{4\Delta}.
\end{align*}
Note that, in contrast to $E_{0}(x)$, the potentials $E_{\pm}(x)$
have imaginary part due to the presence of the excited state $\left\vert e\right\rangle $
in the wave functions of the bright states. This leads to the decay
of the corresponding Bloch states in the bright bands already on the
level of the diagonal Hamiltonians ${\cal H}_{\pm}$, in contrast
to the Bloch states for the dark state with Hermitian ${\cal H}_{0}$.
The potentials $E_{\pm}(x)$ are the standard optical lattice potentials,
giving rise to the standard band structures (see, for example, the
review \cite{bloch2008many} and references therein). The resulting
band structure for the off-resonant case with positive detuning and
$\Gamma=0$ is shown in Fig.~\ref{fig:Figure1}.

\section{Bloch bands of the $\Lambda$-system – numerical solution}

\label{app:numerical}

Our numerical analysis of the system is based on the Hamiltonian (\ref{H1})
written in the basis $|g_{1}\rangle,|g_{2}\rangle,|e\rangle$ and
uses the Bloch ansatz $\psi_{q}(x)=\mathbf{u}(x)e^{iqx}$. Here $\mathbf{u}(x)=(u_{g_{1}}(x),u_{e}(x),u_{g_{2}}(x))^{T}$
is a periodic function with period $\lambda$ {[}the periodicity of
the Hamiltonian (\ref{H1}){]}, and $q\in[-{\pi}/{\lambda},{\pi}/{\lambda}]$
is the quasimomentum.

With this ansatz, the Schrödinger equation for the quasiperiodic Bloch
function $\psi(x)$ turns into an equation for a periodic function
$\mathbf{u}(x)$: 
\begin{widetext}
	\begin{equation}
	\left[\frac{1}{2m}\left(-i\hbar\partial_{x}+q\right)^{2}+\left(\begin{array}{ccc}
	0 & \Omega_{c}(x)/2 & 0\\
	\Omega_{c}(x)/2 & -\widetilde{\Delta} & \Omega_{p}/2\\
	0 & \Omega_{p}/2 & 0
	\end{array}\right)\right]\mathbf{u}(x)=\epsilon(q)\mathbf{u}(x)\label{eqn:E1}
	\end{equation}
	with $q$ being an external parameter. Fourier expansion of the functions
	$u_{a}(x)$ with $a=g_{1,2},e$ gives 
	\begin{equation}
	u_{a}(x)=\sum\limits _{n=-N}^{N}\tilde{u}_{a,n}e^{in\frac{2\pi}{\lambda}x},\label{FT}
	\end{equation}
	which is truncated to $|n|\leq N$, Eq.~(\ref{eqn:E1}) reduces the
	above equation to the matrix eigenvalue problem for a non-Hermitian
	$3(2N+1)\times3(2N+1)$ sparse matrix. 
\end{widetext}

The value $N$ necessary for the convergence of the solution is determined
by the requirement that the expansion (\ref{FT}) correctly represents
rapidly oscillating wave function of the bright channels, which have
visible coupling to the dark channel. For $\epsilon\approx0.1,\Omega_{c},\Delta\in O(10^{4})E_{R}$,
values $N\approx200-400$ are found to be sufficient.

\begin{figure}[!b]
	\includegraphics[clip,width=0.95\columnwidth]{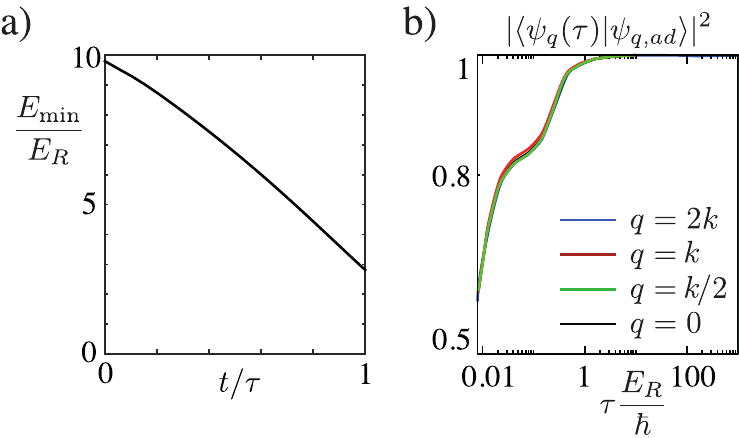} \caption{(Color online) a) The finite gap between the two lowest dark-state
		bands during the whole loading protocol (see text). b) Overlap between
		the final state during the loading protocol $\ket{\psi_{q}(\tau)}$
		%and the target state $\ket{\psi_{q,{ad}}(t)}$ in the lowest dark-state
		Bloch band as a function of the protocol duration $\tau$ for several
		values of the quasimomentum $q$. The chosen parameters of the system
		are $\epsilon=0.1$, $\Omega_{p}=2000E_{R}$, $\Delta=\Gamma=0,V_{0}=30E_{R}$.}
	\label{fig:loading} 
\end{figure}

\section{Loading protocol for lowest Bloch band of the `dark state' channel
	$0$}

\label{app:loading}

In this section we present the details of the protocol for loading
into the lowest Bloch band for the `dark state' channel $0$, referred
to in the main text.

We assume a laser configuration, where the control field is switched
off initially, $\Omega_{c}=0$, while the probe laser $\Omega_{p}=\textrm{const}$,
i.e.~ the dark state is simply $\left\vert g_{1}\right\rangle $).
Initially, we turn on an additional off-resonant optical lattice potential
$V_{L}(x)=V_{0}\cos^{2}(k_{L}x)$ acting on the ground states. Note
that this potential is chosen so that peaks of the periodic potential
$V_{L}(x)$ at positions $n\lambda_{L}/2$ ($n=0,\pm1,\ldots$) match
those of $V_{na}(x)$ (we consider $k_{L}=k$). The protocol consists
of adiabatically turning off the OL $V_{L}(x)$, while turning on
$\Omega_{c}$ to its final value. The lowest Bloch band in the OL
$V_{L}(x)$ is thus mapped to the lowest Bloch band of $V_{na}(x)$,
and an atom prepared in the lowest Bloch band of $V_{L}(x)$ will
be adiabatically transferred to the lowest band of $V_{na}(x)$. Adiabaticity
during the transfer in a time period $\tau$ is guaranteed by the
finite excitation gap during this process, as indicated in Fig.~\ref{fig:loading}a
(and similarly for gaps between the dark and the bright states). To
be specific we choose a linear ramp $V_{0}(t)=V_{0}\cdotp(1-t/\tau)$
and $\Omega_{c}(t)=\Omega_{c}\cdot t/\tau$, where $V_{0}=30E_{R}$.

As an initial state we choose the state $\ket{\psi_{q}}$ with the
quasimomentum $q$ in the lowest Bloch band of the potential $V_{L}(x)$.
The evolution of this state $\ket{\psi_{q}(t)}$ during the protocol
is calculated using the complete Hamiltonian of the system, and Fig.
\ref{fig:loading}b shows the results for the overlap (fidelity) of
the final state of this evolution $\ket{\psi_{q}(\tau)}$ with the
target state $\ket{\psi_{q,{ad}}(t)}$ in the lowest dark-state Bloch
band in the non-adiabatic potential $V_{na}(x)$ for several values
of $q$ as a function of $\tau$. We see that the fidelity approaches
unity for $\tau$ being already few inverse recoil frequencies $\hbar/E_{R}$.
For shorter $\tau$, the fidelity rapidly decreases first when $\hbar/\tau$
becomes of the order of the gap between the first dark-state Bloch
bands ($\sim E_{R}$), and then of the order of the gap to the bright
states ($\sim\Omega_{p}$).

In conclusion, an efficient transfer protocol exists to prepare atoms
in the lowest Bloch band of the `dark state' channel $0$.

\section{Effects of couplings between Born-Oppenheimer channels}

\label{app:coupling}

Here we present analytic considerations of the decay of the dark-state
BO channel due to non-adiabatic couplings to the bright-state channels,
supporting our numerical findings shown in Figs. 2d and 2a in the
main text. We limit the discussion to the lowest dark-state Bloch
band.

\begin{figure}[h]
	\includegraphics[width=8.7cm]{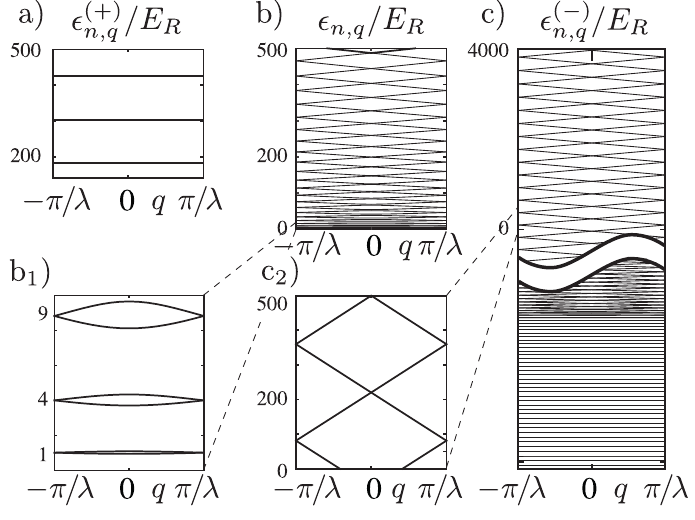} \caption{Band structure for uncoupled BO dark and bright channels. Panel a)
		shows lowest levels for BO channel $+$, a ladder of (for low energies)
		harmonic levels tightly bound in the BO potential $E_{+}(x)$. Panel
		b) band structure for $0$ channel, for BO potential $V_{na}$ with
		accompanying magnification in ${\rm b_{2})}$ of lowest few, gapped
		bands. Panel c) shows extent of bands for the BO channel $-$, from
		the minimum of BO $-$ potential $E_{-}(x).$ through potential threshold,
		up to high-above-threshold regime {[}shown also in the accompanying
		magnification ${\rm c_{2})}${]} where the particle is in almost freely-moving
		with a high-momentum, with a large slope of dispersion relation $\epsilon_{n,q}^{(-)}$
		in ${\rm c_{2}).}$ Parameters: $\Omega_{c}=\Delta=1.7\times10^{4}E_{R}$,
		$\epsilon=0.16$, and $\Gamma=0$.}
	\label{fig:Figure1} 
\end{figure}

The dominant coupling of the dark state to the bright ones is given
by Eq.~(\ref{eq:HC1}). Being proportional to $\alpha^{\prime}(x)$
which is anti-periodic with $\pi/k$, these terms couple quasimomenta
$q$ and $q+k$ (or to $q-k=q+k-2k$). The second order correction
to the energy of the lowest dark state with the quasimomentum $q$
is 
\[
\delta\epsilon_{1,q}=\sum_{\sigma=\pm,n_{\sigma}}\frac{\left\vert M_{\sigma n}(q)\right\vert ^{2}}{\epsilon_{1,q}(q)-\epsilon_{n_{\sigma},q+k}^{(\sigma)}},
\]
where $\sigma=\pm$ refers to the upper and lower bright states, $\epsilon_{n_{\sigma},q}^{(\sigma)}$
are the corresponding dispersions for the $n_{\sigma}$-th band, and
$M_{\sigma n}(q)$ are the coupling matrix elements 
\begin{equation}
\begin{aligned}M_{\sigma n}(q) & =\frac{\hbar^{2}}{2m}\int_{-\pi/2k}^{\pi/2k}dx\alpha^{\prime}(x)N_{\sigma}(x)\\
& \left\{ -\psi_{1,q}(x)\partial_{x}\psi_{n,q+k}^{(\sigma)\ast}(x)+\psi_{n,q+k}^{(\sigma)\ast}(x)\partial_{x}\psi_{1,q}(x)\right\} ,
\end{aligned}
\label{eq:matrix element}
\end{equation}
where $\psi_{n,q}^{(\sigma)}(x)$ and $\psi_{1,q}(x)$ are the wave
functions of the bright $\sigma n$ and lowest dark {[}see Eq.~(\ref{eq:wave function dark}){]}
states, respectively, and we performed integration by part in the
first term.

The imaginary part of $\delta\epsilon_{1,q}$ is 
\begin{align}
\textrm{Im}\,\delta\epsilon_{1,q} & =\sum_{n_{+}}\frac{\left\vert M_{+n}(q)\right\vert ^{2}\textrm{Im}\,\epsilon_{n,q+n}^{(+)}}{[\epsilon_{1,q}-\textrm{Re}\,\epsilon_{n,q+k}^{(+)}]^{2}+[\textrm{Im}\,\epsilon_{n,q+k}^{(+)}]^{2}}\label{Implus}\\
& +\sum_{n_{-}}\frac{\left\vert M_{-n}(q)\right\vert ^{2}\textrm{Im}\,\epsilon_{n,q+k}^{(-)}}{[\epsilon_{1,q}-\textrm{Re}\,\epsilon_{n,q+k}^{(-)}]^{2}+[\textrm{Im}\,\epsilon_{n,q+k}^{(-)}]^{2}},\label{Imminus}
\end{align}
where we write explicitly the contributions from the upper ($+$)
and the lower ($-$) bright states. Keeping in mind that both $\epsilon_{1,q}$
and $\textrm{Im}\,\epsilon_{n,q+k}^{(+)}$ are much smaller than the
energy gap to the upper bright state, the contribution $\textrm{Im}\,\delta\epsilon_{1,q+}$
from the upper bright states can be written as 
\[
\textrm{Im}\,\delta\epsilon_{1,q}{}_{+}\approx\sum_{n_{+}}\left\vert M_{+n}(q)\right\vert ^{2}\frac{\textrm{Im}\,\epsilon_{n,q+k}^{(+)}}{[\textrm{Re}\,\epsilon_{n,q+k}^{(+)}]^{2}}.
\]
The dominant terms come from the lowest bands which are well-described
in the tight-binding approximation by using the localized states $\phi_{n}(x)$
in local potential wells. The coupling matrix element (\ref{eq:matrix element})
involves the first derivative and, as a result, the part $\psi_{\pi}$
of the dark state wave function {[}see Eq.~(\ref{wave function parts}){]}
couples to the even bands, while $\psi_{0}$ to the odd ones. One
can see that the combination in the bracket in Eq.~(\ref{eq:matrix element})
pushes the zeroes of $\phi_{n}(x)$ outside the center of the well
for even $n$, and towards the center for odd $n$. For this reason
the couplings to even $n$ are larger, and keeping them as the sole
contribution we get 
\[
\textrm{Im}\,\delta\epsilon_{1,q}{}_{+}\approx\sin^{2}(\frac{\pi q}{2k})\sum_{\mathrm{even}\,n_{+}}\left\vert m_{+n}^{\pi}\right\vert ^{2}\frac{\textrm{Im}\,\epsilon_{n,q+k}^{(+)}}{[\textrm{Re}\,\epsilon_{n,q+k}^{(+)}]^{2}},
\]
where 
\[
\begin{aligned}m_{+n}^{\pi} & =\frac{\hbar^{2}}{2m}\int_{-\pi/2k}^{\pi/2k}dx\alpha^{\prime}(x)N_{+}(x)\\
& \left\{ -\psi_{\pi}(x)\partial_{x}\phi_{n}(x)+\phi_{n}(x)\partial_{x}\psi_{\pi}(x)\right\} .
\end{aligned}
\]
The $q$-dependence of the dispersion $\epsilon_{n,q+k}^{(+)}$ of
the lowest bands can also be neglected, $\epsilon_{n,q}^{(+)}\approx\epsilon_{n}^{(+)}$
, and we obtain 
\begin{align*}
\textrm{Im}\,\delta\epsilon_{1,q}{}_{+} & \approx\sin^{2}\left(\frac{\pi q}{2k}\right)\sum_{\mathrm{even}\,n_{+}}\left\vert m_{+n}^{\pi}\right\vert ^{2}\frac{\textrm{Im}\,\epsilon_{n}^{(+)}}{[\textrm{Re}\,\epsilon_{n}^{(+)}]^{2}}\\
& \equiv-\frac{1}{2}\sin^{2}\left(\frac{\pi q}{2k}\right)\gamma_{1}^{(0\to+)}\left(\epsilon,\frac{\Omega_{p}}{E_{R}},\frac{\Delta}{\Omega_{p}},\frac{\Gamma}{\Omega_{p}}\right).
\end{align*}

\begin{figure}
	\includegraphics[width=0.95\columnwidth]{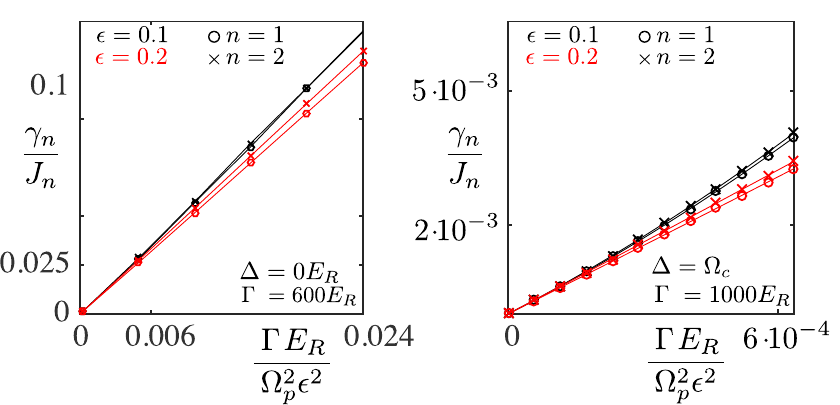} \caption{(Color online) Functions $\gamma_{n}$ determining the regular decay
		of the two lowest Bloch bands $n=1,2$ in the dark-state channel $0$
		for the resonant (left panel) and off-resonant (right panel) cases.}
	\label{fig:Figure2} 
\end{figure}

The contribution {[}see Eq.~(\ref{Imminus}){]} from the lower bright
state has two different parts: A resonant contribution $\textrm{Im}\,\delta\epsilon_{1,q-\mathrm{res}}$
from the band $n_{0}$ with the states, which are resonant to the
dark state for some resonant quasimomentum $q_{\ast}$, $\textrm{Re}\,\epsilon_{n_{0},q_{\ast}+k}^{(-)}\approx\epsilon_{1,q_{\ast}}$,
and a regular one $\textrm{Im}\,\delta\epsilon_{1,q-\mathrm{reg}}$
from the other bands with $\textrm{Re}\,\epsilon_{n_{0},q+k}^{(-)}$
being far from $\epsilon_{1,q}$. The dominant contributions to $\textrm{Im}\,\delta\epsilon_{1,q-\mathrm{reg}}$
come from the bands with energies close to the top of the optical
potential $E_{-}(x)$, for which the wave functions have substantial
amplitudes and oscillate slowly in the coupling region $\alpha^{\prime}(x)\neq0$.
In this region, the wave functions of the relevant states depend only
weakly on $q$, such that the leading $q$-dependence of the coupling
matrix elements is again determined by the coupling to the dark-state
wave function $\psi_{\pi}(x)$, $M_{-n}(q)\approx\sin(\pi q/2k)M_{-n}$.
The decay rates $\textrm{Im}\,\epsilon_{n,q+n}^{(-)}$ of the relevant
bright states can be considered as $q$-independent, $\textrm{Im}\,\epsilon_{n,q+k}^{(-)}\approx\textrm{Im}\,\epsilon_{n}^{(-)}$.
On the other hand, the $q$-dependence of $\textrm{Re}\,\epsilon_{n,q}^{(-)}$
cannot be ignored, but the widths of $\textrm{Re}\,\epsilon_{n,q}^{(-)}$
for the relevant bands are much smaller than the gap ($\approx\min\left\vert E_{-}(x)\right\vert $)
between them and the dark state. As a result, the energy denominator
in the expression for $\textrm{Im}\,\delta\epsilon_{1,q-\mathrm{reg}}$
can be considered as $q$-independent, and $\textrm{Im}\,\delta\epsilon_{1,q-\mathrm{res}}$,
therefore, has the same form as $\textrm{Im}\,\delta\epsilon_{1,q+\mathrm{reg}}$
with a different function $\gamma_{1}^{(0\to-)}(\epsilon,\Omega_{p}/E_{R},\Delta/\Omega_{p},\Gamma/\Omega_{p})$.
The sum of the two regular contributions from the $+$ and $-$ channels
give the function $\gamma_{1}$ mentioned in the main text, Eq.~(3).
Similar considerations for the higher Bloch bands in the dark-state
channel give that the dominant $q$-dependence of the decay rate originates
from the antisymmetric part of the Bloch wave function (\ref{eq:Bloch wf n}),
and we obtain $\gamma_{n,q}=\gamma_{n}\cos^{2}(\pi q/2k-\pi n/2)$.
In Figs.~\ref{fig:Figure2}a and \ref{fig:Figure2}b we show the
numerical results for the dependence of the functions $\gamma_{1}$
and $\gamma_{2}$ on the parameters of the system, both for the resonant
and the off-resonant cases. We find out that the functions $\gamma_{n}$
depend almost linearly on the parameter $\Gamma E_{R}^{2}/\Omega_{p}^{2}\epsilon$.

In calculating the resonant contribution $\delta\epsilon_{1,q-\mathrm{res}}=\delta\epsilon_{1,q\mathrm{res}}$,
we can take into account the semiclassical character of the band $n_{0}$
and write $\epsilon_{1,q}-\textrm{Re}\,\epsilon_{n_{0},q+k}^{(-)}\approx v(q-q_{\ast})$
for $q$ close to $q_{\ast}$ (see Fig.~2e in the main text). Here
$v$ is the group velocity (slope of the band $n_{0}$) at $q=q_{\ast}$.
Keeping in mind that $M_{-n_{0}}(q)$ and $\textrm{Im}\,\epsilon_{n_{0},q}^{(-)}$
are slow functions of $q$, we can write the resonant contribution
in the form 
\[
\delta\epsilon_{1,q\mathrm{res}}=\left\vert M_{-n_{0}}(q_{\ast})\right\vert ^{2}\frac{1}{v(q-q_{\ast})+i\Gamma_{\ast}/2},
\]
where $\Gamma_{\ast}=-2\textrm{Im}\,\epsilon_{n_{0},q_{\ast}+k}^{(-)}$.
The imaginary part of this expression has the typical resonant Lorentzian
structure with the width proportional to $\Gamma_{\ast}$, and the
height scales with $\Gamma_{\ast}^{-1}$. This structure is, therefore,
visible if $\Gamma_{\ast}$ is much smaller than the bandwidth $B_{n_{0}}\sim vk$
of the band $n_{0}$ (Fig.~2e in the main text). Another condition
is related to the strength of the coupling $M_{-n_{0}}(q_{\ast})$.
This matrix element is visible only for zero and negative detunings,
and is exponentially small for positive detuning when the wave function
of the bright state in the band $n_{0}$ strongly oscillates in the
coupling regions. The resonant contribution is therefore invisible
in the latter case. For the resonant case with $\Gamma_{\ast}=\Gamma/4$
and $B_{n_{0}}\sim\sqrt{\Omega_{c}E_{R}}$, the visibility condition
reads $\Gamma/\Omega_{p}\ll\sqrt{\epsilon\kappa}$, while for the
negatively-detuned off-resonant case with $\Gamma_{\ast}\approx(\Omega_{p}/2\Delta)^{2}\Gamma$
and $B_{n_{0}}\sim\Omega_{c}\sqrt{E_{R}/\left\vert \Delta\right\vert }$,
the resonances appear if $\Gamma/\left\vert \Delta\right\vert \ll\sqrt{\kappa}$.

To conclude, the discussion of the decay of the dark-state Bloch bands,
we compare our results with those of the Kronig-Penney model for the
periodic set of $\delta$-functional potentials $\widetilde{V}_{\delta}(x)=(\hbar^{2}A/2m)\delta(x)$
which strength has a small (negative) imaginary part, $A=\pi k/2\epsilon-iB$
with $0<B\ll\pi k/2\epsilon$, mimicking the decay due to the coupling
to other channels. The simplest way to obtain the energy spectrum
in this case is by using the analytic continuation to the complex
interaction strength in Eq. (\ref{eq:epsilon for delta}) with the
result for the decay rate in the lowest band 
\[
\gamma_{1,q}^{(\delta)}=-\frac{2}{\hbar}\textrm{Im}\,\epsilon_{1,q}^{(\delta)}\approx\frac{E_{R}}{\hbar}\frac{32\epsilon^{2}B}{\pi^{3}k}\left(1+\cos\left(\frac{\pi q}{k}\right)\right)\sim\cos^{2}\frac{\pi q}{2k},
\]
which, in contrast to the expression (3) from the main text, is maximal
in the center of the Brillouin zone.

\begin{figure}[!t]
	\includegraphics[width=8.5cm]{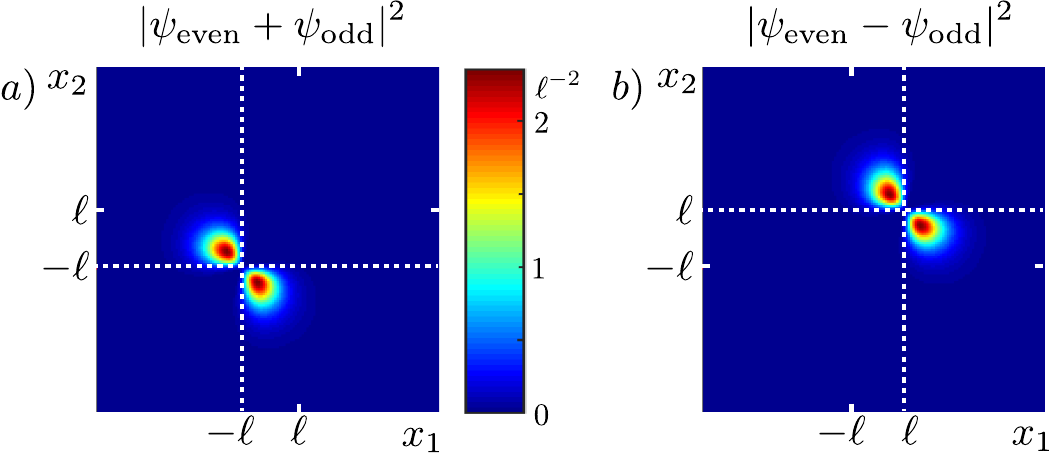} \caption{(Color online) The modulus-square of the sum a) and of the difference
		b) of the wave functions of the two lowest two-particles eigenstates
		{[}even $\psi_{\textrm{even}}(x_{1},x_{2})$ and odd $\psi_{\textrm{odd}}(x_{1},x_{2})${]}
		of the Hamiltonian (4) in the main text for $a_{D}=6.3l$. The white
		dashed lines mark boundaries of the subwavelength domain, where the
		local dipole moment $d(x)$ changes its sign.}
	
	\label{fig:boundstates} 
\end{figure}

\section{Domain wall molecules - details of numerical analysis}

\label{app:domainwall}

The wave functions and the eigenenergies for the domain-wall molecules
shown in Fig.~3 of the main text, are obtained by the numerical diagonalization
of the two-particle Hamiltonian (4) of the main text with $V_{L}(x)=0$.
We performed our calculations for the case $d_{1}=-d_{2}=d$ such
that the position-dependent dipolar moment of the dark state $d(x)$
crosses zero at $x=\pm\ell$ {[}$d(x)<0$ for $\left|x\right|<\ell${]},
see Fig.~3a of the main text. In our calculations we limit the coordinates
of the particles to the intervals $\left|x_{1,2}\right|\leq6\ell$
with zero boundary condition for the wave function. This region is
then discretized into a uniform grid of dimension $500\times500$.
We also assume strong harmonic confinement for the motion in the transverse
directions, such that the particles occupy only the lowest transverse
Gaussian modes $\phi_{0}(\xi=y,z)=\exp(-\xi^{2}/2\ell_{\perp}^{2})/\sqrt{2\pi^{1/2}\ell_{\perp}}$,
with $\ell_{\perp}\ll\ell$. Under this condition, the effective interparticle
interaction $V(x_{1},\!x_{2})$ is obtained by projecting the 3D dipole-dipole
interaction onto the lowest transverse Gaussian modes, %\begin{widetext}

\[
\begin{aligned}V(x_{1},\!x_{2})\! & =\!d(x_{1})d(x_{2})\int\limits _{\mathbb{R}^{4}}\!\!\frac{\left|\vec{r}_{12}\right|^{2}-3z_{12}^{2}}{\left|\vec{r}_{12}\right|^{5}}\prod\limits _{i=1}^{2}\phi_{0}^{2}(y_{i})\phi_{0}^{2}(z_{i}){\rm d}y_{i}{\rm d}z_{i},\\
& =d(x_{1})d(x_{2})\frac{F(x_{12}/\sqrt{2}\ell_{\perp})}{2\sqrt{2}\ell_{\perp}^{3}}\\
& \to\frac{d(x_{1})d(x_{2})}{\left|x_{12}\right|^{3}},\qquad\left|x_{12}\right|\gg\ell_{\perp},
\end{aligned}
\]
where we assume dipoles oriented along the $z$-axis, $\vec{r}_{12}=\vec{r}_{1}-\vec{r}_{2}$,
and $F(s)=\sqrt{\pi}(2s^{2}+1)\exp(s^{2})[1-\textrm{erf}\,(s)]-2s$
with $\textrm{erf}\,(s)$ being the error function. We neglect here
the contact term in the pseudopotential for the quasi-1D scattering
with dipole-dipole interaction \cite{You2000,Bohn2006,Santos2007},
as well as the short-range part of the interparticle interaction.
The effect of these term is negligible because the wave function of
the bound state becomes very small when two particles approach each
other, see Fig. \ref{fig:boundstates}. This results from the strong
repulsive interaction when the two particles are close on the same
side of the interface. On the other hand, when they approach each
other from different sides, the attractive interaction between them
vanishes. This makes such configurations unfavorable for the bound
state. Note that effective interaction $V(x_{1},\!x_{2})$ depends
on $l_{\perp}$. This dependence, however, manifests itself only for
$\left|x_{12}\right|\lesssim\ell_{\perp}\ll\ell$, and practically
does not affect the results of our calculations (for $\ell_{\perp}\lesssim\ell/6$)
because the wave functions for the states of interest are very small
in this region (see Figs.~\ref{fig:boundstates}a and b).

The lowest two eigenstates are found to be symmetric (even) and antisymmetric
(odd) superpositions of the two states in which the molecular wave
function is located around $x=-\ell$ or $x=\ell$ (dashed circles
in Fig.~3a of the main text and Fig.~\ref{fig:boundstates}), with
the corresponding wave functions $\psi_{\textrm{even}}(x_{1},x_{2})$
and $\psi_{\textrm{odd}}(x_{1},x_{2})$. To demonstrate the structure
of these states we plot the modulus squared of the sum (Fig.~\ref{fig:boundstates}a)
and of the difference (Fig.~\ref{fig:boundstates}b) of their wave
functions for the case $a_{D}=6.3l$. The dependence of the eigenenergies
of these states on the strength of the interparticle interaction is
presented in Fig.~3a of the main text. The energy difference $\Delta E$
between the two eigenstates provides the molecular hopping element,
$J_{{\rm pair}}=\Delta E/2$ (see Fig.~3c of the main text), between
the domain-wall molecular states at $x=-\ell$ (Fig.~\ref{fig:boundstates}a)
and $x=\ell$ (Fig.~\ref{fig:boundstates}a).

We also performed the analogous calculations for the case of three
particles and found the formation (for $a_{d}\gtrsim5\ell$) of the
three-body bound state (trimer) in which one particle is in the region
$\left|x\right|<\ell$ and the other two are on the opposite sides
of this region. The formation of the trimer can be intuitively understood
by considering the interaction of the two particles forming a domain
molecule, say at the interface $x=-\ell$ , with the third particle
in the region $x>\ell$. It is easy to see that this interaction is
attractive, thus giving rise to the formation of the bound state.
This also explains why the threshold for the formation of the trimer
is slightly lower than that ($a_{d}/\ell\ge 6$) for the dimer (molecule).

\begin{figure}[!t]
	\includegraphics[width=8.5cm]{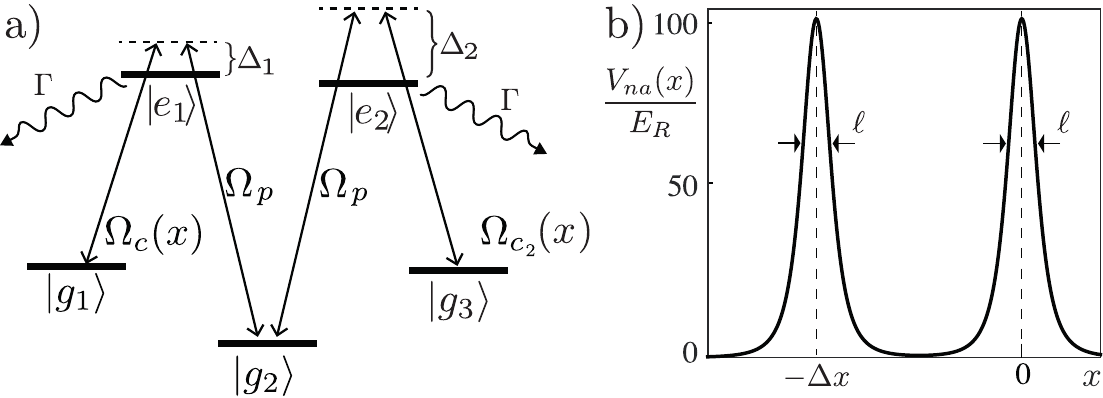}\caption{Panel a) Double-$\Lambda$ configuration with two position-dependent
		Rabi frequencies of the control fields $\Omega_{c_{1}}(x)$, $\Omega_{c_{2}}(x)$
		and two homogeneous Rabi frequencies $\Omega_{p_{1}}$, $\Omega_{p_{2}}$
		of the probe fields (see text). b) Dark-state non-adiabatic potential
		$V_{na}(x)$in units of the recoil energy $E_{R}$ for the case $\Omega_{c_{1}}=\Omega_{c_{2}}$,
		$\Omega_{p_{1}}=\Omega_{p_{2}}$, and $\epsilon_{1}=\epsilon_{2}=0.1$. }
	
	\label{fig:twoLambda} 
\end{figure}

\section{Double-peaked subwavelength optical barriers }

\label{app:doublelambda}

We present here another example for a laser-atom configuration as
an atomic double-$\Lambda$ configuration, which leads to the double-peaked
structure in the non-adiabatic potential of the dark state. This is
similar to that presented in Fig.~4 of the main text, but with the
possibility of controlling the spatial separation between the peaks.
In contrast to the atomic scheme described in the main text, we consider
here three ground states $\ket{g_{1}}$, $\ket{g_{2}}$, and $\ket{g_{3}}$,
and two excited states $\ket{e_{1}}$ and $\ket{e_{2}}$, which might
belong to Zeeman manifolds with different angular momentum $J$ and,
therefore, the laser couplings between them are generated by two independent
control and two independent probe lasers. The lasers are tuned to
satisfy the Raman on-resonance condition between the ground states
$\ket{g_{1}}$ and $\ket{g_{2}}$, as well as between $\ket{g_{2}}$
and $\ket{g_{3}}$, with the constant Rabi frequencies $\Omega_{p_{1}}$
and $\Omega_{p_{2}}$for the probe beams, and position dependent Rabi
frequencies $\Omega_{c_{1}}(x)=\Omega_{c_{1}}\sin(kx)$ and $\Omega_{c_{2}}(x)=\Omega_{c_{2}}\sin[k(x+\Delta x)]$
(with off-set $\Delta x$) for the control beams (see Fig.~\ref{fig:twoLambda}a).
For a weak probe beams, $\epsilon_{i}=\Omega_{p_{i}}/\Omega_{c_{i}}\ll1$,
the calculated non-adiabatic potential for the dark state $\ket{E_{0}}\sim\Omega_{p_{1}}\Omega_{c_{2}}(x)\ket{g_{1}}-\Omega_{c_{1}}(x)\Omega_{c_{2}}(x)\ket{g_{2}}+\Omega_{p_{2}}\Omega_{c_{1}}(x)\ket{g_{3}}$
contains two peaks of widths $\ell_{i}\sim\epsilon_{i}\lambda/2\pi=\epsilon_{i}/k$:
the first one originates from the left leg of the double-$\Lambda$
system, and a second one from the right leg. These are separated by
the distance $\Delta x$ (we assume $\Delta x>\ell_{i}$). The separation
of the two barriers can, therefore, be controlled by tuning the relative
phase $k\Delta x$ between the control beams. Fig.~\ref{fig:twoLambda}b
shows the resulting non-adiabatic potential $V_{na}(x)$ in units
of the recoil energy $E_{R}=\hbar^{2}k^{2}/2m$ for the case $\Omega_{p_{1}}=\Omega_{p_{2}}$
and $\epsilon_{1}=\epsilon_{2}=0.1$.

\section{Measurement proposals}

\label{app:Measurements}

Here we briefly discuss possible measurement schemes for testing the
results of the paper related to both single-atom physics and two-atom/many-atom
physics.

On the level of single-atom physics (i.e. both for the 1D bandstructure
and, in a similar way, for the double wire/layer) time of flight would
be a natural way to map the spatial structure to momentum structures
\cite{Gerbier2005,Bloch2005,bloch2008many}. The new aspect here is
the correlation (as given by the spatial variation of the dark state)
between the atomic internal states and subwavelength structures. 

Another possibility would be to reverse in time our loading protocole
(see Section \ref{app:loading}), so that subwavelength structures
are mapped to band fillings etc. in standard optical lattices, where
well-developed tools exist for read out \cite{Bloch2005}. Finally,
one can modulate the ‘dark state’ lattice parameters, and perform
a spectroscopy by driving transitions between bands (see for example\cite{Stoferle2004,Kollath2006,bloch2008many}).
On the level of two and three atoms, the same strategy results in
the mapping of the two- and three-body bound states to occupations
of atomic vibrational levels in the lattice $V_{L}(x)$. Another way
to perform spectroscopy of the bound pairs and trimers would be driving
with an external field (e.g. RF \cite{Gupta2003,Shin2007}) which
couples to the atomic internal states (analog to the coupling to spin-degree
of freedom in the case of atomic Cooper pairs \cite{Torma2000,Chin2004}
and Feshbach molecules \cite{Regal2003}), or to use a lattice modulation
spectroscopy.

We plan to include the details of all the topics discussed in the
Supplemental Materials in Ref. \cite{Lacki2016inp}.


\begin{thebibliography}{57}%
	\makeatletter
	\providecommand \@ifxundefined [1]{%
		\@ifx{#1\undefined}
	}%
	\providecommand \@ifnum [1]{%
		\ifnum #1\expandafter \@firstoftwo
		\else \expandafter \@secondoftwo
		\fi
	}%
	\providecommand \@ifx [1]{%
		\ifx #1\expandafter \@firstoftwo
		\else \expandafter \@secondoftwo
		\fi
	}%
	\providecommand \natexlab [1]{#1}%
	\providecommand \enquote  [1]{``#1''}%
	\providecommand \bibnamefont  [1]{#1}%
	\providecommand \bibfnamefont [1]{#1}%
	\providecommand \citenamefont [1]{#1}%
	\providecommand \href@noop [0]{\@secondoftwo}%
	\providecommand \href [0]{\begingroup \@sanitize@url \@href}%
	\providecommand \@href[1]{\@@startlink{#1}\@@href}%
	\providecommand \@@href[1]{\endgroup#1\@@endlink}%
	\providecommand \@sanitize@url [0]{\catcode `\\12\catcode `\$12\catcode
		`\&12\catcode `\#12\catcode `\^12\catcode `\_12\catcode `\%12\relax}%
	\providecommand \@@startlink[1]{}%
	\providecommand \@@endlink[0]{}%
	\providecommand \url  [0]{\begingroup\@sanitize@url \@url }%
	\providecommand \@url [1]{\endgroup\@href {#1}{\urlprefix }}%
	\providecommand \urlprefix  [0]{URL }%
	\providecommand \Eprint [0]{\href }%
	\providecommand \doibase [0]{http://dx.doi.org/}%
	\providecommand \selectlanguage [0]{\@gobble}%
	\providecommand \bibinfo  [0]{\@secondoftwo}%
	\providecommand \bibfield  [0]{\@secondoftwo}%
	\providecommand \translation [1]{[#1]}%
	\providecommand \BibitemOpen [0]{}%
	\providecommand \bibitemStop [0]{}%
	\providecommand \bibitemNoStop [0]{.\EOS\space}%
	\providecommand \EOS [0]{\spacefactor3000\relax}%
	\providecommand \BibitemShut  [1]{\csname bibitem#1\endcsname}%
	\let\auto@bib@innerbib\@empty
	%</preamble>
	\bibitem [{\citenamefont {Gu{\'e}ry-Odelin}\ and\ \citenamefont
		{Cohen-Tannoudji}(2011)}]{Cohen-Tannoudji2011}%
	\BibitemOpen
	\bibfield  {author} {\bibinfo {author} {\bibfnamefont {D.}~\bibnamefont
			{Gu{\'e}ry-Odelin}}\ and\ \bibinfo {author} {\bibfnamefont {C.}~\bibnamefont
			{Cohen-Tannoudji}},\ }\href@noop {} {\emph {\bibinfo {title} {Advances in
				Atomic Physics}}}\ (\bibinfo  {publisher} {World Scientific},\ \bibinfo
	{year} {2011})\BibitemShut {NoStop}%
	\bibitem [{\citenamefont {Lewenstein}\ \emph {et~al.}(2012)\citenamefont
		{Lewenstein}, \citenamefont {Sanpera},\ and\ \citenamefont
		{Ahufinger}}]{Lewenstein2012}%
	\BibitemOpen
	\bibfield  {author} {\bibinfo {author} {\bibfnamefont {M.}~\bibnamefont
			{Lewenstein}}, \bibinfo {author} {\bibfnamefont {A.}~\bibnamefont {Sanpera}},
		\ and\ \bibinfo {author} {\bibfnamefont {V.}~\bibnamefont {Ahufinger}},\
	}\href@noop {} {\emph {\bibinfo {title} {Ultracold Atoms in Optical Lattices:
			Simulating quantum many-body systems}}}\ (\bibinfo  {publisher} {OUP
	Oxford},\ \bibinfo {year} {2012})\BibitemShut {NoStop}%
\bibitem [{\citenamefont {Bloch}\ \emph {et~al.}(2008)\citenamefont {Bloch},
	\citenamefont {Dalibard},\ and\ \citenamefont {Zwerger}}]{bloch2008many}%
\BibitemOpen
\bibfield  {author} {\bibinfo {author} {\bibfnamefont {I.}~\bibnamefont
		{Bloch}}, \bibinfo {author} {\bibfnamefont {J.}~\bibnamefont {Dalibard}}, \
	and\ \bibinfo {author} {\bibfnamefont {W.}~\bibnamefont {Zwerger}},\
}\href@noop {} {\bibfield  {journal} {\bibinfo  {journal} {Rev. Mod. Phys.}\
}\textbf {\bibinfo {volume} {80}},\ \bibinfo {pages} {885} (\bibinfo {year}
{2008})}\BibitemShut {NoStop}%
\bibitem [{\citenamefont {Mitsch}\ \emph {et~al.}(2014)\citenamefont {Mitsch},
	\citenamefont {Sayrin}, \citenamefont {Albrecht}, \citenamefont
	{Schneeweiss},\ and\ \citenamefont {Rauschenbeutel}}]{Rauschenbeutel2014}%
\BibitemOpen
\bibfield  {author} {\bibinfo {author} {\bibfnamefont {R.}~\bibnamefont
		{Mitsch}}, \bibinfo {author} {\bibfnamefont {C.}~\bibnamefont {Sayrin}},
	\bibinfo {author} {\bibfnamefont {B.}~\bibnamefont {Albrecht}}, \bibinfo
	{author} {\bibfnamefont {P.}~\bibnamefont {Schneeweiss}}, \ and\ \bibinfo
	{author} {\bibfnamefont {A.}~\bibnamefont {Rauschenbeutel}},\ }\href@noop {}
{\bibfield  {journal} {\bibinfo  {journal} {Nat. Commun.}\ }\textbf {\bibinfo
		{volume} {5}} (\bibinfo {year} {2014})}\BibitemShut {NoStop}%
\bibitem [{\citenamefont {Thompson}\ \emph {et~al.}(2013)\citenamefont
	{Thompson}, \citenamefont {Tiecke}, \citenamefont {de~Leon}, \citenamefont
	{Feist}, \citenamefont {Akimov}, \citenamefont {Gullans}, \citenamefont
	{Zibrov}, \citenamefont {Vuleti{\'c}},\ and\ \citenamefont
	{Lukin}}]{Thompson2013}%
\BibitemOpen
\bibfield  {author} {\bibinfo {author} {\bibfnamefont {J.}~\bibnamefont
		{Thompson}}, \bibinfo {author} {\bibfnamefont {T.}~\bibnamefont {Tiecke}},
	\bibinfo {author} {\bibfnamefont {N.}~\bibnamefont {de~Leon}}, \bibinfo
	{author} {\bibfnamefont {J.}~\bibnamefont {Feist}}, \bibinfo {author}
	{\bibfnamefont {A.}~\bibnamefont {Akimov}}, \bibinfo {author} {\bibfnamefont
		{M.}~\bibnamefont {Gullans}}, \bibinfo {author} {\bibfnamefont
		{A.}~\bibnamefont {Zibrov}}, \bibinfo {author} {\bibfnamefont
		{V.}~\bibnamefont {Vuleti{\'c}}}, \ and\ \bibinfo {author} {\bibfnamefont
		{M.}~\bibnamefont {Lukin}},\ }\href@noop {} {\bibfield  {journal} {\bibinfo
		{journal} {Science}\ }\textbf {\bibinfo {volume} {340}},\ \bibinfo {pages}
	{1202} (\bibinfo {year} {2013})}\BibitemShut {NoStop}%
\bibitem [{\citenamefont {Gonz{\'a}lez-Tudela}\ \emph
	{et~al.}(2015)\citenamefont {Gonz{\'a}lez-Tudela}, \citenamefont {Hung},
	\citenamefont {Chang}, \citenamefont {Cirac},\ and\ \citenamefont
	{Kimble}}]{Kimble2015}%
\BibitemOpen
\bibfield  {author} {\bibinfo {author} {\bibfnamefont {A.}~\bibnamefont
		{Gonz{\'a}lez-Tudela}}, \bibinfo {author} {\bibfnamefont {C.-L.}\
		\bibnamefont {Hung}}, \bibinfo {author} {\bibfnamefont {D.~E.}\ \bibnamefont
		{Chang}}, \bibinfo {author} {\bibfnamefont {J.~I.}\ \bibnamefont {Cirac}}, \
	and\ \bibinfo {author} {\bibfnamefont {H.}~\bibnamefont {Kimble}},\
}\href@noop {} {\bibfield  {journal} {\bibinfo  {journal} {Nature Photon.}\
}\textbf {\bibinfo {volume} {9}},\ \bibinfo {pages} {320} (\bibinfo {year}
{2015})}\BibitemShut {NoStop}%
\bibitem [{\citenamefont {Chang}\ \emph {et~al.}(2009)\citenamefont {Chang},
	\citenamefont {Thompson}, \citenamefont {Park}, \citenamefont {Vuleti{\'c}},
	\citenamefont {Zibrov}, \citenamefont {Zoller},\ and\ \citenamefont
	{Lukin}}]{Lukin2009}%
\BibitemOpen
\bibfield  {author} {\bibinfo {author} {\bibfnamefont {D.~E.}\ \bibnamefont
		{Chang}}, \bibinfo {author} {\bibfnamefont {J.~D.}\ \bibnamefont {Thompson}},
	\bibinfo {author} {\bibfnamefont {H.}~\bibnamefont {Park}}, \bibinfo {author}
	{\bibfnamefont {V.}~\bibnamefont {Vuleti{\'c}}}, \bibinfo {author}
	{\bibfnamefont {A.~S.}\ \bibnamefont {Zibrov}}, \bibinfo {author}
	{\bibfnamefont {P.}~\bibnamefont {Zoller}}, \ and\ \bibinfo {author}
	{\bibfnamefont {M.~D.}\ \bibnamefont {Lukin}},\ }\href@noop {} {\bibfield
	{journal} {\bibinfo  {journal} {Phys. Rev. Lett.}\ }\textbf {\bibinfo
		{volume} {103}},\ \bibinfo {pages} {123004} (\bibinfo {year}
	{2009})}\BibitemShut {NoStop}%
\bibitem [{\citenamefont {Gullans}\ \emph {et~al.}(2012)\citenamefont
	{Gullans}, \citenamefont {Tiecke}, \citenamefont {Chang}, \citenamefont
	{Feist}, \citenamefont {Thompson}, \citenamefont {Cirac}, \citenamefont
	{Zoller},\ and\ \citenamefont {Lukin}}]{Lukin2012}%
\BibitemOpen
\bibfield  {author} {\bibinfo {author} {\bibfnamefont {M.}~\bibnamefont
		{Gullans}}, \bibinfo {author} {\bibfnamefont {T.~G.}\ \bibnamefont {Tiecke}},
	\bibinfo {author} {\bibfnamefont {D.~E.}\ \bibnamefont {Chang}}, \bibinfo
	{author} {\bibfnamefont {J.}~\bibnamefont {Feist}}, \bibinfo {author}
	{\bibfnamefont {J.~D.}\ \bibnamefont {Thompson}}, \bibinfo {author}
	{\bibfnamefont {J.~I.}\ \bibnamefont {Cirac}}, \bibinfo {author}
	{\bibfnamefont {P.}~\bibnamefont {Zoller}}, \ and\ \bibinfo {author}
	{\bibfnamefont {M.~D.}\ \bibnamefont {Lukin}},\ }\href@noop {} {\bibfield
	{journal} {\bibinfo  {journal} {Phys. Rev. Lett.}\ }\textbf {\bibinfo
		{volume} {109}},\ \bibinfo {pages} {235309} (\bibinfo {year}
	{2012})}\BibitemShut {NoStop}%
\bibitem [{\citenamefont {Yi}\ \emph {et~al.}(2008)\citenamefont {Yi},
	\citenamefont {Daley}, \citenamefont {Pupillo},\ and\ \citenamefont
	{Zoller}}]{yi2008state}%
\BibitemOpen
\bibfield  {author} {\bibinfo {author} {\bibfnamefont {W.}~\bibnamefont
		{Yi}}, \bibinfo {author} {\bibfnamefont {A.}~\bibnamefont {Daley}}, \bibinfo
	{author} {\bibfnamefont {G.}~\bibnamefont {Pupillo}}, \ and\ \bibinfo
	{author} {\bibfnamefont {P.}~\bibnamefont {Zoller}},\ }\href@noop {}
{\bibfield  {journal} {\bibinfo  {journal} {New J. Phys.}\ }\textbf {\bibinfo
		{volume} {10}},\ \bibinfo {pages} {073015} (\bibinfo {year}
	{2008})}\BibitemShut {NoStop}%
\bibitem [{\citenamefont {Kiffner}\ \emph {et~al.}(2008)\citenamefont
	{Kiffner}, \citenamefont {Evers},\ and\ \citenamefont
	{Zubairy}}]{Zubairy2008}%
\BibitemOpen
\bibfield  {author} {\bibinfo {author} {\bibfnamefont {M.}~\bibnamefont
		{Kiffner}}, \bibinfo {author} {\bibfnamefont {J.}~\bibnamefont {Evers}}, \
	and\ \bibinfo {author} {\bibfnamefont {M.~S.}\ \bibnamefont {Zubairy}},\
}\href@noop {} {\bibfield  {journal} {\bibinfo  {journal} {Phys. Rev. Lett.}\
}\textbf {\bibinfo {volume} {100}},\ \bibinfo {pages} {073602} (\bibinfo
{year} {2008})}\BibitemShut {NoStop}%
\bibitem [{\citenamefont {Nascimbene}\ \emph {et~al.}(2015)\citenamefont
	{Nascimbene}, \citenamefont {Goldman}, \citenamefont {Cooper},\ and\
	\citenamefont {Dalibard}}]{nascimbene2015dynamic}%
\BibitemOpen
\bibfield  {author} {\bibinfo {author} {\bibfnamefont {S.}~\bibnamefont
		{Nascimbene}}, \bibinfo {author} {\bibfnamefont {N.}~\bibnamefont {Goldman}},
	\bibinfo {author} {\bibfnamefont {N.~R.}\ \bibnamefont {Cooper}}, \ and\
	\bibinfo {author} {\bibfnamefont {J.}~\bibnamefont {Dalibard}},\ }\href@noop
{} {\bibfield  {journal} {\bibinfo  {journal} {Phys. Rev. Lett.}\ }\textbf
	{\bibinfo {volume} {115}},\ \bibinfo {pages} {140401} (\bibinfo {year}
	{2015})}\BibitemShut {NoStop}%
\bibitem [{\citenamefont {Ritt}\ \emph {et~al.}(2006)\citenamefont {Ritt},
	\citenamefont {Geckeler}, \citenamefont {Salger}, \citenamefont {Cennini},\
	and\ \citenamefont {Weitz}}]{Ritt:2006cb}%
\BibitemOpen
\bibfield  {author} {\bibinfo {author} {\bibfnamefont {G.}~\bibnamefont
		{Ritt}}, \bibinfo {author} {\bibfnamefont {C.}~\bibnamefont {Geckeler}},
	\bibinfo {author} {\bibfnamefont {T.}~\bibnamefont {Salger}}, \bibinfo
	{author} {\bibfnamefont {G.}~\bibnamefont {Cennini}}, \ and\ \bibinfo
	{author} {\bibfnamefont {M.}~\bibnamefont {Weitz}},\ }\href@noop {}
{\bibfield  {journal} {\bibinfo  {journal} {Phys. Rev. A}\ }\textbf {\bibinfo
		{volume} {74}},\ \bibinfo {pages} {063622} (\bibinfo {year}
	{2006})}\BibitemShut {NoStop}%
\bibitem [{\citenamefont {Brezger}\ \emph {et~al.}(1999)\citenamefont
	{Brezger}, \citenamefont {Schulze}, \citenamefont {Schmidt}, \citenamefont
	{Pfau},\ and\ \citenamefont {Mlynek}}]{brezger1999polarization}%
\BibitemOpen
\bibfield  {author} {\bibinfo {author} {\bibfnamefont {B.}~\bibnamefont
		{Brezger}}, \bibinfo {author} {\bibfnamefont {T.}~\bibnamefont {Schulze}},
	\bibinfo {author} {\bibfnamefont {P.}~\bibnamefont {Schmidt}}, \bibinfo
	{author} {\bibfnamefont {R.~M.~T.}\ \bibnamefont {Pfau}}, \ and\ \bibinfo
	{author} {\bibfnamefont {J.}~\bibnamefont {Mlynek}},\ }\href@noop {}
{\bibfield  {journal} {\bibinfo  {journal} {Europhys. Lett.}\ }\textbf
	{\bibinfo {volume} {46}},\ \bibinfo {pages} {148} (\bibinfo {year}
	{1999})}\BibitemShut {NoStop}%
\bibitem [{\citenamefont {Salger}\ \emph {et~al.}(2007)\citenamefont {Salger},
	\citenamefont {Geckeler}, \citenamefont {Kling},\ and\ \citenamefont
	{Weitz}}]{Salger2007}%
\BibitemOpen
\bibfield  {author} {\bibinfo {author} {\bibfnamefont {T.}~\bibnamefont
		{Salger}}, \bibinfo {author} {\bibfnamefont {C.}~\bibnamefont {Geckeler}},
	\bibinfo {author} {\bibfnamefont {S.}~\bibnamefont {Kling}}, \ and\ \bibinfo
	{author} {\bibfnamefont {M.}~\bibnamefont {Weitz}},\ }\href@noop {}
{\bibfield  {journal} {\bibinfo  {journal} {Phys. Rev. Lett.}\ }\textbf
	{\bibinfo {volume} {99}},\ \bibinfo {pages} {190405} (\bibinfo {year}
	{2007})}\BibitemShut {NoStop}%
\bibitem [{\citenamefont {Lundblad}\ \emph {et~al.}(2008)\citenamefont
	{Lundblad}, \citenamefont {Lee}, \citenamefont {Spielman}, \citenamefont
	{Brown}, \citenamefont {Phillips},\ and\ \citenamefont
	{Porto}}]{lundblad2008atoms}%
\BibitemOpen
\bibfield  {author} {\bibinfo {author} {\bibfnamefont {N.}~\bibnamefont
		{Lundblad}}, \bibinfo {author} {\bibfnamefont {P.~J.}\ \bibnamefont {Lee}},
	\bibinfo {author} {\bibfnamefont {I.~B.}\ \bibnamefont {Spielman}}, \bibinfo
	{author} {\bibfnamefont {B.~L.}\ \bibnamefont {Brown}}, \bibinfo {author}
	{\bibfnamefont {W.~D.}\ \bibnamefont {Phillips}}, \ and\ \bibinfo {author}
	{\bibfnamefont {J.~V.}\ \bibnamefont {Porto}},\ }\href@noop {} {\bibfield
	{journal} {\bibinfo  {journal} {Phys. Rev. Lett.}\ }\textbf {\bibinfo
		{volume} {100}},\ \bibinfo {pages} {150401} (\bibinfo {year}
	{2008})}\BibitemShut {NoStop}%
\bibitem [{\citenamefont {Lukin}(2003)}]{lukin2003colloquium}%
\BibitemOpen
\bibfield  {author} {\bibinfo {author} {\bibfnamefont {M.}~\bibnamefont
		{Lukin}},\ }\href@noop {} {\bibfield  {journal} {\bibinfo  {journal} {Rev.
			Mod. Phys.}\ }\textbf {\bibinfo {volume} {75}},\ \bibinfo {pages} {457}
	(\bibinfo {year} {2003})}\BibitemShut {NoStop}%
\bibitem [{\citenamefont {Vitanov}\ \emph {et~al.}(2016)\citenamefont
	{Vitanov}, \citenamefont {Rangelov}, \citenamefont {Shore},\ and\
	\citenamefont {Bergmann}}]{vitanov2016stimulated}%
\BibitemOpen
\bibfield  {author} {\bibinfo {author} {\bibfnamefont {N.~V.}\ \bibnamefont
		{Vitanov}}, \bibinfo {author} {\bibfnamefont {A.~A.}\ \bibnamefont
		{Rangelov}}, \bibinfo {author} {\bibfnamefont {B.~W.}\ \bibnamefont {Shore}},
	\ and\ \bibinfo {author} {\bibfnamefont {K.}~\bibnamefont {Bergmann}},\
}\href@noop {} {\bibfield  {journal} {\bibinfo  {journal} {arXiv preprint
		arXiv:1605.00224}\ } (\bibinfo {year} {2016})}\BibitemShut {NoStop}%
\bibitem [{\citenamefont {Baier}\ \emph {et~al.}(2016)\citenamefont {Baier},
	\citenamefont {Mark}, \citenamefont {Petter}, \citenamefont {Aikawa},
	\citenamefont {Chomaz}, \citenamefont {Cai}, \citenamefont {Baranov},
	\citenamefont {Zoller},\ and\ \citenamefont {Ferlaino}}]{Ferlaino2016}%
\BibitemOpen
\bibfield  {author} {\bibinfo {author} {\bibfnamefont {S.}~\bibnamefont
		{Baier}}, \bibinfo {author} {\bibfnamefont {M.}~\bibnamefont {Mark}},
	\bibinfo {author} {\bibfnamefont {D.}~\bibnamefont {Petter}}, \bibinfo
	{author} {\bibfnamefont {K.}~\bibnamefont {Aikawa}}, \bibinfo {author}
	{\bibfnamefont {L.}~\bibnamefont {Chomaz}}, \bibinfo {author} {\bibfnamefont
		{Z.}~\bibnamefont {Cai}}, \bibinfo {author} {\bibfnamefont {M.}~\bibnamefont
		{Baranov}}, \bibinfo {author} {\bibfnamefont {P.}~\bibnamefont {Zoller}}, \
	and\ \bibinfo {author} {\bibfnamefont {F.}~\bibnamefont {Ferlaino}},\
}\href@noop {} {\bibfield  {journal} {\bibinfo  {journal} {Science}\ }\textbf
{\bibinfo {volume} {352}},\ \bibinfo {pages} {201} (\bibinfo {year}
{2016})}\BibitemShut {NoStop}%
\bibitem [{\citenamefont {Burdick}\ \emph {et~al.}(2016)\citenamefont
	{Burdick}, \citenamefont {Tang},\ and\ \citenamefont {Lev}}]{Lev2016}%
\BibitemOpen
\bibfield  {author} {\bibinfo {author} {\bibfnamefont {N.~Q.}\ \bibnamefont
		{Burdick}}, \bibinfo {author} {\bibfnamefont {Y.}~\bibnamefont {Tang}}, \
	and\ \bibinfo {author} {\bibfnamefont {B.~L.}\ \bibnamefont {Lev}},\
}\href@noop {} {\bibfield  {journal} {\bibinfo  {journal} {arXiv:1605.03211}\
} (\bibinfo {year} {2016})}\BibitemShut {NoStop}%
\bibitem [{\citenamefont {Kadau}\ \emph {et~al.}(2016)\citenamefont {Kadau},
	\citenamefont {Schmitt}, \citenamefont {Wenzel}, \citenamefont {Wink},
	\citenamefont {Maier}, \citenamefont {Ferrier-Barbut},\ and\ \citenamefont
	{Pfau}}]{Kadau:2016cb}%
\BibitemOpen
\bibfield  {author} {\bibinfo {author} {\bibfnamefont {H.}~\bibnamefont
		{Kadau}}, \bibinfo {author} {\bibfnamefont {M.}~\bibnamefont {Schmitt}},
	\bibinfo {author} {\bibfnamefont {M.}~\bibnamefont {Wenzel}}, \bibinfo
	{author} {\bibfnamefont {C.}~\bibnamefont {Wink}}, \bibinfo {author}
	{\bibfnamefont {T.}~\bibnamefont {Maier}}, \bibinfo {author} {\bibfnamefont
		{I.}~\bibnamefont {Ferrier-Barbut}}, \ and\ \bibinfo {author} {\bibfnamefont
		{T.}~\bibnamefont {Pfau}},\ }\href@noop {} {\bibfield  {journal} {\bibinfo
		{journal} {Nature}\ }\textbf {\bibinfo {volume} {530}},\ \bibinfo {pages}
	{194} (\bibinfo {year} {2016})}\BibitemShut {NoStop}%
\bibitem [{\citenamefont {Ferrier-Barbut}\ \emph {et~al.}(2016)\citenamefont
	{Ferrier-Barbut}, \citenamefont {Kadau}, \citenamefont {Schmitt},
	\citenamefont {Wenzel},\ and\ \citenamefont {Pfau}}]{FerrierBarbut:2016jo}%
\BibitemOpen
\bibfield  {author} {\bibinfo {author} {\bibfnamefont {I.}~\bibnamefont
		{Ferrier-Barbut}}, \bibinfo {author} {\bibfnamefont {H.}~\bibnamefont
		{Kadau}}, \bibinfo {author} {\bibfnamefont {M.}~\bibnamefont {Schmitt}},
	\bibinfo {author} {\bibfnamefont {M.}~\bibnamefont {Wenzel}}, \ and\ \bibinfo
	{author} {\bibfnamefont {T.}~\bibnamefont {Pfau}},\ }\href@noop {} {\bibfield
	{journal} {\bibinfo  {journal} {Phys. Rev. Lett.}\ }\textbf {\bibinfo
		{volume} {116}},\ \bibinfo {pages} {215301} (\bibinfo {year}
	{2016})}\BibitemShut {NoStop}%
\bibitem [{\citenamefont {de~Paz}\ \emph {et~al.}(2013)\citenamefont {de~Paz},
	\citenamefont {Sharma}, \citenamefont {Chotia}, \citenamefont {Mar{\'e}chal},
	\citenamefont {Huckans}, \citenamefont {Pedri}, \citenamefont {Santos},
	\citenamefont {Gorceix}, \citenamefont {Vernac},\ and\ \citenamefont
	{Laburthe-Tolra}}]{dePaz:2013ff}%
\BibitemOpen
\bibfield  {author} {\bibinfo {author} {\bibfnamefont {A.}~\bibnamefont
		{de~Paz}}, \bibinfo {author} {\bibfnamefont {A.}~\bibnamefont {Sharma}},
	\bibinfo {author} {\bibfnamefont {A.}~\bibnamefont {Chotia}}, \bibinfo
	{author} {\bibfnamefont {E.}~\bibnamefont {Mar{\'e}chal}}, \bibinfo {author}
	{\bibfnamefont {J.~H.}\ \bibnamefont {Huckans}}, \bibinfo {author}
	{\bibfnamefont {P.}~\bibnamefont {Pedri}}, \bibinfo {author} {\bibfnamefont
		{L.}~\bibnamefont {Santos}}, \bibinfo {author} {\bibfnamefont
		{O.}~\bibnamefont {Gorceix}}, \bibinfo {author} {\bibfnamefont
		{L.}~\bibnamefont {Vernac}}, \ and\ \bibinfo {author} {\bibfnamefont
		{B.}~\bibnamefont {Laburthe-Tolra}},\ }\href@noop {} {\bibfield  {journal}
	{\bibinfo  {journal} {Phys. Rev. Lett.}\ }\textbf {\bibinfo {volume} {111}},\
	\bibinfo {pages} {185305} (\bibinfo {year} {2013})}\BibitemShut {NoStop}%
\bibitem [{\citenamefont {Gardiner}\ and\ \citenamefont
	{Zoller}(2015)}]{gardiner2015quantum}%
\BibitemOpen
\bibfield  {author} {\bibinfo {author} {\bibfnamefont {C.}~\bibnamefont
		{Gardiner}}\ and\ \bibinfo {author} {\bibfnamefont {P.}~\bibnamefont
		{Zoller}},\ }in\ \href@noop {} {\emph {\bibinfo {booktitle} {The Quantum
			World of Ultra-Cold Atoms and Light Book II: The Physics of Quantum-Optical
			Devices}}}\ (\bibinfo  {publisher} {World Scientific},\ \bibinfo {year}
{2015})\ pp.\ \bibinfo {pages} {1--524}\BibitemShut {NoStop}%
\bibitem [{\citenamefont {Gorshkov}\ \emph {et~al.}(2008)\citenamefont
	{Gorshkov}, \citenamefont {Jiang}, \citenamefont {Greiner}, \citenamefont
	{Zoller},\ and\ \citenamefont {Lukin}}]{gorshkov2008coherent}%
\BibitemOpen
\bibfield  {author} {\bibinfo {author} {\bibfnamefont {A.~V.}\ \bibnamefont
		{Gorshkov}}, \bibinfo {author} {\bibfnamefont {L.}~\bibnamefont {Jiang}},
	\bibinfo {author} {\bibfnamefont {M.}~\bibnamefont {Greiner}}, \bibinfo
	{author} {\bibfnamefont {P.}~\bibnamefont {Zoller}}, \ and\ \bibinfo {author}
	{\bibfnamefont {M.~D.}\ \bibnamefont {Lukin}},\ }\href@noop {} {\bibfield
	{journal} {\bibinfo  {journal} {Phys. Rev. Lett.}\ }\textbf {\bibinfo
		{volume} {100}},\ \bibinfo {pages} {093005} (\bibinfo {year}
	{2008})}\BibitemShut {NoStop}%
\bibitem [{Note1()}]{Note1}%
\BibitemOpen
\bibinfo {note} {This is in contrast to spin-orbit coupling schemes based on
	running wave laser configurations with $\Lambda $-systems \cite
	{Spielman2013}}\BibitemShut {NoStop}%
\bibitem [{\citenamefont {Kazantsev}\ \emph {et~al.}(1990)\citenamefont
	{Kazantsev}, \citenamefont {Surdutovich},\ and\ \citenamefont
	{Yakovlev}}]{kazantsev1990mechanical}%
\BibitemOpen
\bibfield  {author} {\bibinfo {author} {\bibfnamefont {A.~P.}\ \bibnamefont
		{Kazantsev}}, \bibinfo {author} {\bibfnamefont {G.}~\bibnamefont
		{Surdutovich}}, \ and\ \bibinfo {author} {\bibfnamefont {V.}~\bibnamefont
		{Yakovlev}},\ }\href@noop {} {\emph {\bibinfo {title} {Mechanical action of
			light on atoms}}}\ (\bibinfo  {publisher} {World Scientific},\ \bibinfo
{year} {1990})\BibitemShut {NoStop}%
\bibitem [{\citenamefont {Dum}\ and\ \citenamefont
	{Olshanii}(1996)}]{Olshanii1996}%
\BibitemOpen
\bibfield  {author} {\bibinfo {author} {\bibfnamefont {R.}~\bibnamefont
		{Dum}}\ and\ \bibinfo {author} {\bibfnamefont {M.}~\bibnamefont {Olshanii}},\
}\href {\doibase 10.1103/PhysRevLett.76.1788} {\bibfield  {journal} {\bibinfo
	{journal} {Phys. Rev. Lett.}\ }\textbf {\bibinfo {volume} {76}},\ \bibinfo
{pages} {1788} (\bibinfo {year} {1996})}\BibitemShut {NoStop}%
\bibitem [{\citenamefont {Dutta}\ \emph {et~al.}(1999)\citenamefont {Dutta},
	\citenamefont {Teo},\ and\ \citenamefont {Raithel}}]{Dutta1999}%
\BibitemOpen
\bibfield  {author} {\bibinfo {author} {\bibfnamefont {S.~K.}\ \bibnamefont
		{Dutta}}, \bibinfo {author} {\bibfnamefont {B.~K.}\ \bibnamefont {Teo}}, \
	and\ \bibinfo {author} {\bibfnamefont {G.}~\bibnamefont {Raithel}},\
}\href@noop {} {\bibfield  {journal} {\bibinfo  {journal} {Phys. Rev. Lett.}\
}\textbf {\bibinfo {volume} {83}},\ \bibinfo {pages} {1934} (\bibinfo {year}
{1999})}\BibitemShut {NoStop}%
\bibitem [{\citenamefont {Ruseckas}\ \emph {et~al.}(2005)\citenamefont
	{Ruseckas}, \citenamefont {Juzeli\ifmmode~\bar{u}\else \={u}\fi{}nas},
	\citenamefont {\"Ohberg},\ and\ \citenamefont {Fleischhauer}}]{Ruseckas2005}%
\BibitemOpen
\bibfield  {author} {\bibinfo {author} {\bibfnamefont {J.}~\bibnamefont
		{Ruseckas}}, \bibinfo {author} {\bibfnamefont {G.}~\bibnamefont
		{Juzeli\ifmmode~\bar{u}\else \={u}\fi{}nas}}, \bibinfo {author}
	{\bibfnamefont {P.}~\bibnamefont {\"Ohberg}}, \ and\ \bibinfo {author}
	{\bibfnamefont {M.}~\bibnamefont {Fleischhauer}},\ }\href {\doibase
	10.1103/PhysRevLett.95.010404} {\bibfield  {journal} {\bibinfo  {journal}
		{Phys. Rev. Lett.}\ }\textbf {\bibinfo {volume} {95}},\ \bibinfo {pages}
	{010404} (\bibinfo {year} {2005})}\BibitemShut {NoStop}%
\bibitem [{\citenamefont {Bergmann}\ \emph {et~al.}(2015)\citenamefont
	{Bergmann}, \citenamefont {Vitanov},\ and\ \citenamefont
	{Shore}}]{bergmann2015perspective}%
\BibitemOpen
\bibfield  {author} {\bibinfo {author} {\bibfnamefont {K.}~\bibnamefont
		{Bergmann}}, \bibinfo {author} {\bibfnamefont {N.~V.}\ \bibnamefont
		{Vitanov}}, \ and\ \bibinfo {author} {\bibfnamefont {B.~W.}\ \bibnamefont
		{Shore}},\ }\href@noop {} {\bibfield  {journal} {\bibinfo  {journal} {J.
			Chem. Phys.}\ }\textbf {\bibinfo {volume} {142}},\ \bibinfo {pages} {170901}
	(\bibinfo {year} {2015})}\BibitemShut {NoStop}%
\bibitem [{\citenamefont {Merzbacher}(1998)}]{Merzbacher1998}%
\BibitemOpen
\bibfield  {author} {\bibinfo {author} {\bibfnamefont {E.}~\bibnamefont
		{Merzbacher}},\ }\href@noop {} {\emph {\bibinfo {title} {Quantum
			Mechanics}}}\ (\bibinfo  {publisher} {Wiley},\ \bibinfo {year}
{1998})\BibitemShut {NoStop}%
\bibitem [{SM()}]{SM}%
\BibitemOpen
\href@noop {} {}\bibinfo {note} {{See Appendices}}\BibitemShut {NoStop}%
\bibitem [{\citenamefont {Wang}\ \emph {et~al.}(2006)\citenamefont {Wang},
	\citenamefont {Lukin},\ and\ \citenamefont {Demler}}]{Wang:2006ja}%
\BibitemOpen
\bibfield  {author} {\bibinfo {author} {\bibfnamefont {D.-W.}\ \bibnamefont
		{Wang}}, \bibinfo {author} {\bibfnamefont {M.~D.}\ \bibnamefont {Lukin}}, \
	and\ \bibinfo {author} {\bibfnamefont {E.}~\bibnamefont {Demler}},\
}\href@noop {} {\bibfield  {journal} {\bibinfo  {journal} {Phys. Rev. Lett.}\
}\textbf {\bibinfo {volume} {97}},\ \bibinfo {pages} {180413} (\bibinfo
{year} {2006})}\BibitemShut {NoStop}%
\bibitem [{\citenamefont {Pikovski}\ \emph {et~al.}(2010)\citenamefont
	{Pikovski}, \citenamefont {Klawunn}, \citenamefont {Shlyapnikov},\ and\
	\citenamefont {Santos}}]{Santos2010}%
\BibitemOpen
\bibfield  {author} {\bibinfo {author} {\bibfnamefont {A.}~\bibnamefont
		{Pikovski}}, \bibinfo {author} {\bibfnamefont {M.}~\bibnamefont {Klawunn}},
	\bibinfo {author} {\bibfnamefont {G.~V.}\ \bibnamefont {Shlyapnikov}}, \ and\
	\bibinfo {author} {\bibfnamefont {L.}~\bibnamefont {Santos}},\ }\href@noop {}
{\bibfield  {journal} {\bibinfo  {journal} {Phys. Rev. Lett.}\ }\textbf
	{\bibinfo {volume} {105}},\ \bibinfo {pages} {215302} (\bibinfo {year}
	{2010})}\BibitemShut {NoStop}%
\bibitem [{\citenamefont {Baranov}\ \emph {et~al.}(2011)\citenamefont
	{Baranov}, \citenamefont {Micheli}, \citenamefont {Ronen},\ and\
	\citenamefont {Zoller}}]{Zoller2011}%
\BibitemOpen
\bibfield  {author} {\bibinfo {author} {\bibfnamefont {M.~A.}\ \bibnamefont
		{Baranov}}, \bibinfo {author} {\bibfnamefont {A.}~\bibnamefont {Micheli}},
	\bibinfo {author} {\bibfnamefont {S.}~\bibnamefont {Ronen}}, \ and\ \bibinfo
	{author} {\bibfnamefont {P.}~\bibnamefont {Zoller}},\ }\href@noop {}
{\bibfield  {journal} {\bibinfo  {journal} {Phys. Rev. A}\ }\textbf {\bibinfo
		{volume} {83}},\ \bibinfo {pages} {043602} (\bibinfo {year}
	{2011})}\BibitemShut {NoStop}%
\bibitem [{\citenamefont {Baranov}\ \emph {et~al.}(2012)\citenamefont
	{Baranov}, \citenamefont {Dalmonte}, \citenamefont {Pupillo},\ and\
	\citenamefont {Zoller}}]{Baranov:ChemRev}%
\BibitemOpen
\bibfield  {author} {\bibinfo {author} {\bibfnamefont {M.~A.}\ \bibnamefont
		{Baranov}}, \bibinfo {author} {\bibfnamefont {M.}~\bibnamefont {Dalmonte}},
	\bibinfo {author} {\bibfnamefont {G.}~\bibnamefont {Pupillo}}, \ and\
	\bibinfo {author} {\bibfnamefont {P.}~\bibnamefont {Zoller}},\ }\href@noop {}
{\bibfield  {journal} {\bibinfo  {journal} {Chem. Rev.}\ }\textbf {\bibinfo
		{volume} {112}},\ \bibinfo {pages} {5012} (\bibinfo {year}
	{2012})}\BibitemShut {NoStop}%
\bibitem [{\citenamefont {Yan}\ \emph {et~al.}(2013)\citenamefont {Yan},
	\citenamefont {Moses}, \citenamefont {Gadway}, \citenamefont {Covey},
	\citenamefont {Hazzard}, \citenamefont {Rey}, \citenamefont {Jin},\ and\
	\citenamefont {Ye}}]{Ye2013}%
\BibitemOpen
\bibfield  {author} {\bibinfo {author} {\bibfnamefont {B.}~\bibnamefont
		{Yan}}, \bibinfo {author} {\bibfnamefont {S.~A.}\ \bibnamefont {Moses}},
	\bibinfo {author} {\bibfnamefont {B.}~\bibnamefont {Gadway}}, \bibinfo
	{author} {\bibfnamefont {J.~P.}\ \bibnamefont {Covey}}, \bibinfo {author}
	{\bibfnamefont {K.~R.}\ \bibnamefont {Hazzard}}, \bibinfo {author}
	{\bibfnamefont {A.~M.}\ \bibnamefont {Rey}}, \bibinfo {author} {\bibfnamefont
		{D.~S.}\ \bibnamefont {Jin}}, \ and\ \bibinfo {author} {\bibfnamefont
		{J.}~\bibnamefont {Ye}},\ }\href@noop {} {\bibfield  {journal} {\bibinfo
		{journal} {Nature}\ }\textbf {\bibinfo {volume} {501}},\ \bibinfo {pages}
	{521} (\bibinfo {year} {2013})}\BibitemShut {NoStop}%
\bibitem [{\citenamefont {Moses}\ \emph {et~al.}(2015)\citenamefont {Moses},
	\citenamefont {Covey}, \citenamefont {Miecnikowski}, \citenamefont {Yan},
	\citenamefont {Gadway}, \citenamefont {Ye},\ and\ \citenamefont
	{Jin}}]{Jin2015}%
\BibitemOpen
\bibfield  {author} {\bibinfo {author} {\bibfnamefont {S.~A.}\ \bibnamefont
		{Moses}}, \bibinfo {author} {\bibfnamefont {J.~P.}\ \bibnamefont {Covey}},
	\bibinfo {author} {\bibfnamefont {M.~T.}\ \bibnamefont {Miecnikowski}},
	\bibinfo {author} {\bibfnamefont {B.}~\bibnamefont {Yan}}, \bibinfo {author}
	{\bibfnamefont {B.}~\bibnamefont {Gadway}}, \bibinfo {author} {\bibfnamefont
		{J.}~\bibnamefont {Ye}}, \ and\ \bibinfo {author} {\bibfnamefont {D.~S.}\
		\bibnamefont {Jin}},\ }\href@noop {} {\bibfield  {journal} {\bibinfo
		{journal} {Science}\ }\textbf {\bibinfo {volume} {350}},\ \bibinfo {pages}
	{659} (\bibinfo {year} {2015})}\BibitemShut {NoStop}%
\bibitem [{foo()}]{footb}%
\BibitemOpen
\href@noop {} {}\bibinfo {note} {{For transverse confinement $\ell_{\perp}$
		with a non-resonant optical lattice we note the condition
		$\ell>\ell_{\perp}$.}}\BibitemShut {Stop}%
\bibitem [{\citenamefont {Mandel}\ \emph {et~al.}(2003)\citenamefont {Mandel},
	\citenamefont {Greiner}, \citenamefont {Widera}, \citenamefont {Rom},
	\citenamefont {H{\"a}nsch},\ and\ \citenamefont {Bloch}}]{Mandel:2003ej}%
\BibitemOpen
\bibfield  {author} {\bibinfo {author} {\bibfnamefont {O.}~\bibnamefont
		{Mandel}}, \bibinfo {author} {\bibfnamefont {M.}~\bibnamefont {Greiner}},
	\bibinfo {author} {\bibfnamefont {A.}~\bibnamefont {Widera}}, \bibinfo
	{author} {\bibfnamefont {T.}~\bibnamefont {Rom}}, \bibinfo {author}
	{\bibfnamefont {T.~W.}\ \bibnamefont {H{\"a}nsch}}, \ and\ \bibinfo {author}
	{\bibfnamefont {I.}~\bibnamefont {Bloch}},\ }\href@noop {} {\bibfield
	{journal} {\bibinfo  {journal} {Phys. Rev. Lett.}\ }\textbf {\bibinfo
		{volume} {91}},\ \bibinfo {pages} {010407} (\bibinfo {year}
	{2003})}\BibitemShut {NoStop}%
\bibitem [{\citenamefont {Lee}\ \emph {et~al.}(2007)\citenamefont {Lee},
	\citenamefont {Anderlini}, \citenamefont {Brown}, \citenamefont
	{Sebby-Strabley}, \citenamefont {Phillips},\ and\ \citenamefont
	{Porto}}]{Lee:2007ip}%
\BibitemOpen
\bibfield  {author} {\bibinfo {author} {\bibfnamefont {P.~J.}\ \bibnamefont
		{Lee}}, \bibinfo {author} {\bibfnamefont {M.}~\bibnamefont {Anderlini}},
	\bibinfo {author} {\bibfnamefont {B.~L.}\ \bibnamefont {Brown}}, \bibinfo
	{author} {\bibfnamefont {J.}~\bibnamefont {Sebby-Strabley}}, \bibinfo
	{author} {\bibfnamefont {W.~D.}\ \bibnamefont {Phillips}}, \ and\ \bibinfo
	{author} {\bibfnamefont {J.~V.}\ \bibnamefont {Porto}},\ }\href@noop {}
{\bibfield  {journal} {\bibinfo  {journal} {Phys. Rev. Lett.}\ }\textbf
	{\bibinfo {volume} {99}},\ \bibinfo {pages} {020402} (\bibinfo {year}
	{2007})}\BibitemShut {NoStop}%
\bibitem [{\citenamefont {Soltan-Panahi}\ \emph {et~al.}(2011)\citenamefont
	{Soltan-Panahi}, \citenamefont {Struck}, \citenamefont {Hauke}, \citenamefont
	{Bick}, \citenamefont {Plenkers}, \citenamefont {Meineke}, \citenamefont
	{Becker}, \citenamefont {Windpassinger}, \citenamefont {Lewenstein},\ and\
	\citenamefont {Sengstock}}]{SoltanPanahi:2011ey}%
\BibitemOpen
\bibfield  {author} {\bibinfo {author} {\bibfnamefont {P.}~\bibnamefont
		{Soltan-Panahi}}, \bibinfo {author} {\bibfnamefont {J.}~\bibnamefont
		{Struck}}, \bibinfo {author} {\bibfnamefont {P.}~\bibnamefont {Hauke}},
	\bibinfo {author} {\bibfnamefont {A.}~\bibnamefont {Bick}}, \bibinfo {author}
	{\bibfnamefont {W.}~\bibnamefont {Plenkers}}, \bibinfo {author}
	{\bibfnamefont {G.}~\bibnamefont {Meineke}}, \bibinfo {author} {\bibfnamefont
		{C.}~\bibnamefont {Becker}}, \bibinfo {author} {\bibfnamefont
		{P.}~\bibnamefont {Windpassinger}}, \bibinfo {author} {\bibfnamefont
		{M.}~\bibnamefont {Lewenstein}}, \ and\ \bibinfo {author} {\bibfnamefont
		{K.}~\bibnamefont {Sengstock}},\ }\href@noop {} {\bibfield  {journal}
	{\bibinfo  {journal} {Nat. Phys.}\ }\textbf {\bibinfo {volume} {7}},\
	\bibinfo {pages} {434} (\bibinfo {year} {2011})}\BibitemShut {NoStop}%
\bibitem [{\citenamefont {Dai}\ \emph {et~al.}(2016)\citenamefont {Dai},
	\citenamefont {Yang}, \citenamefont {Reingruber}, \citenamefont {Xu},
	\citenamefont {Jiang}, \citenamefont {Chen}, \citenamefont {Yuan},\ and\
	\citenamefont {Pan}}]{Dai:us}%
\BibitemOpen
\bibfield  {author} {\bibinfo {author} {\bibfnamefont {H.-N.}\ \bibnamefont
		{Dai}}, \bibinfo {author} {\bibfnamefont {B.}~\bibnamefont {Yang}}, \bibinfo
	{author} {\bibfnamefont {A.}~\bibnamefont {Reingruber}}, \bibinfo {author}
	{\bibfnamefont {X.-F.}\ \bibnamefont {Xu}}, \bibinfo {author} {\bibfnamefont
		{X.}~\bibnamefont {Jiang}}, \bibinfo {author} {\bibfnamefont {Y.-A.}\
		\bibnamefont {Chen}}, \bibinfo {author} {\bibfnamefont {Z.-S.}\ \bibnamefont
		{Yuan}}, \ and\ \bibinfo {author} {\bibfnamefont {J.-W.}\ \bibnamefont
		{Pan}},\ }\href@noop {} {\bibfield  {journal} {\bibinfo  {journal} {Nat.
			Phys.}\ } (\bibinfo {year} {2016})}\BibitemShut {NoStop}%
\bibitem [{\citenamefont {Lin}\ \emph {et~al.}(2011)\citenamefont {Lin},
	\citenamefont {Jim{\'e}nez-Garc{\'\i}a},\ and\ \citenamefont
	{Spielman}}]{Lin:2011hn}%
\BibitemOpen
\bibfield  {author} {\bibinfo {author} {\bibfnamefont {Y.~J.}\ \bibnamefont
		{Lin}}, \bibinfo {author} {\bibfnamefont {K.}~\bibnamefont
		{Jim{\'e}nez-Garc{\'\i}a}}, \ and\ \bibinfo {author} {\bibfnamefont {I.~B.}\
		\bibnamefont {Spielman}},\ }\href@noop {} {\bibfield  {journal} {\bibinfo
		{journal} {Nature}\ }\textbf {\bibinfo {volume} {471}},\ \bibinfo {pages}
	{83} (\bibinfo {year} {2011})}\BibitemShut {NoStop}%
\bibitem [{\citenamefont {Wang}\ \emph {et~al.}(2012)\citenamefont {Wang},
	\citenamefont {Yu}, \citenamefont {Fu}, \citenamefont {Miao}, \citenamefont
	{Huang}, \citenamefont {Chai}, \citenamefont {Zhai},\ and\ \citenamefont
	{Zhang}}]{Wang:2012gv}%
\BibitemOpen
\bibfield  {author} {\bibinfo {author} {\bibfnamefont {P.}~\bibnamefont
		{Wang}}, \bibinfo {author} {\bibfnamefont {Z.-Q.}\ \bibnamefont {Yu}},
	\bibinfo {author} {\bibfnamefont {Z.}~\bibnamefont {Fu}}, \bibinfo {author}
	{\bibfnamefont {J.}~\bibnamefont {Miao}}, \bibinfo {author} {\bibfnamefont
		{L.}~\bibnamefont {Huang}}, \bibinfo {author} {\bibfnamefont
		{S.}~\bibnamefont {Chai}}, \bibinfo {author} {\bibfnamefont {H.}~\bibnamefont
		{Zhai}}, \ and\ \bibinfo {author} {\bibfnamefont {J.}~\bibnamefont {Zhang}},\
}\href@noop {} {\bibfield  {journal} {\bibinfo  {journal} {Phys. Rev. Lett.}\
}\textbf {\bibinfo {volume} {109}},\ \bibinfo {pages} {095301} (\bibinfo
{year} {2012})}\BibitemShut {NoStop}%
\bibitem [{\citenamefont {Cheuk}\ \emph {et~al.}(2012)\citenamefont {Cheuk},
	\citenamefont {Sommer}, \citenamefont {Hadzibabic}, \citenamefont {Yefsah},
	\citenamefont {Bakr},\ and\ \citenamefont {Zwierlein}}]{Cheuk:2012tl}%
\BibitemOpen
\bibfield  {author} {\bibinfo {author} {\bibfnamefont {L.~W.}\ \bibnamefont
		{Cheuk}}, \bibinfo {author} {\bibfnamefont {A.~T.}\ \bibnamefont {Sommer}},
	\bibinfo {author} {\bibfnamefont {Z.}~\bibnamefont {Hadzibabic}}, \bibinfo
	{author} {\bibfnamefont {T.}~\bibnamefont {Yefsah}}, \bibinfo {author}
	{\bibfnamefont {W.~S.}\ \bibnamefont {Bakr}}, \ and\ \bibinfo {author}
	{\bibfnamefont {M.~W.}\ \bibnamefont {Zwierlein}},\ }\href@noop {} {\bibfield
	{journal} {\bibinfo  {journal} {Phys. Rev. Lett.}\ }\textbf {\bibinfo
		{volume} {109}},\ \bibinfo {pages} {095302} (\bibinfo {year}
	{2012})}\BibitemShut {NoStop}%
\bibitem [{\citenamefont {Stuhl}\ \emph {et~al.}(2015)\citenamefont {Stuhl},
	\citenamefont {Lu}, \citenamefont {Aycock}, \citenamefont {Genkina},\ and\
	\citenamefont {Spielman}}]{Stuhl:2015cb}%
\BibitemOpen
\bibfield  {author} {\bibinfo {author} {\bibfnamefont {B.~K.}\ \bibnamefont
		{Stuhl}}, \bibinfo {author} {\bibfnamefont {H.~I.}\ \bibnamefont {Lu}},
	\bibinfo {author} {\bibfnamefont {L.~M.}\ \bibnamefont {Aycock}}, \bibinfo
	{author} {\bibfnamefont {D.}~\bibnamefont {Genkina}}, \ and\ \bibinfo
	{author} {\bibfnamefont {I.~B.}\ \bibnamefont {Spielman}},\ }\href@noop {}
{\bibfield  {journal} {\bibinfo  {journal} {Science}\ }\textbf {\bibinfo
		{volume} {349}},\ \bibinfo {pages} {1514} (\bibinfo {year}
	{2015})}\BibitemShut {NoStop}%
\bibitem [{\citenamefont {Huang}\ \emph {et~al.}(2016)\citenamefont {Huang},
	\citenamefont {Meng}, \citenamefont {Wang}, \citenamefont {Peng},
	\citenamefont {Zhang}, \citenamefont {Chen}, \citenamefont {Li},
	\citenamefont {Zhou},\ and\ \citenamefont {Zhang}}]{Huang:2016kf}%
\BibitemOpen
\bibfield  {author} {\bibinfo {author} {\bibfnamefont {L.}~\bibnamefont
		{Huang}}, \bibinfo {author} {\bibfnamefont {Z.}~\bibnamefont {Meng}},
	\bibinfo {author} {\bibfnamefont {P.}~\bibnamefont {Wang}}, \bibinfo {author}
	{\bibfnamefont {P.}~\bibnamefont {Peng}}, \bibinfo {author} {\bibfnamefont
		{S.-L.}\ \bibnamefont {Zhang}}, \bibinfo {author} {\bibfnamefont
		{L.}~\bibnamefont {Chen}}, \bibinfo {author} {\bibfnamefont {D.}~\bibnamefont
		{Li}}, \bibinfo {author} {\bibfnamefont {Q.}~\bibnamefont {Zhou}}, \ and\
	\bibinfo {author} {\bibfnamefont {J.}~\bibnamefont {Zhang}},\ }\href@noop {}
{\bibfield  {journal} {\bibinfo  {journal} {Nat. Phys.}\ }\textbf {\bibinfo
		{volume} {12}},\ \bibinfo {pages} {540} (\bibinfo {year} {2016})}\BibitemShut
{NoStop}%
\bibitem [{\citenamefont {Galitski}\ and\ \citenamefont
	{Spielman}(2013)}]{Spielman2013}%
\BibitemOpen
\bibfield  {author} {\bibinfo {author} {\bibfnamefont {V.}~\bibnamefont
		{Galitski}}\ and\ \bibinfo {author} {\bibfnamefont {I.~B.}\ \bibnamefont
		{Spielman}},\ }\href@noop {} {\bibfield  {journal} {\bibinfo  {journal}
		{Nature}\ }\textbf {\bibinfo {volume} {494}},\ \bibinfo {pages} {49}
	(\bibinfo {year} {2013})}\BibitemShut {NoStop}%
\bibitem [{\citenamefont {Dalibard}\ \emph {et~al.}(2011)\citenamefont
	{Dalibard}, \citenamefont {Gerbier}, \citenamefont {Juzeli{\=u}nas},\ and\
	\citenamefont {{\"O}hberg}}]{Dalibard:2011gg}%
\BibitemOpen
\bibfield  {author} {\bibinfo {author} {\bibfnamefont {J.}~\bibnamefont
		{Dalibard}}, \bibinfo {author} {\bibfnamefont {F.}~\bibnamefont {Gerbier}},
	\bibinfo {author} {\bibfnamefont {G.}~\bibnamefont {Juzeli{\=u}nas}}, \ and\
	\bibinfo {author} {\bibfnamefont {P.}~\bibnamefont {{\"O}hberg}},\
}\href@noop {} {\bibfield  {journal} {\bibinfo  {journal} {Rev. Mod. Phys.}\
}\textbf {\bibinfo {volume} {83}},\ \bibinfo {pages} {1523} (\bibinfo {year}
{2011})}\BibitemShut {NoStop}%
\bibitem [{\citenamefont {Goldman}\ \emph {et~al.}(2016)\citenamefont
	{Goldman}, \citenamefont {Budich},\ and\ \citenamefont
	{Zoller}}]{Goldman:2016tw}%
\BibitemOpen
\bibfield  {author} {\bibinfo {author} {\bibfnamefont {N.}~\bibnamefont
		{Goldman}}, \bibinfo {author} {\bibfnamefont {J.~C.}\ \bibnamefont {Budich}},
	\ and\ \bibinfo {author} {\bibfnamefont {P.}~\bibnamefont {Zoller}},\
}\href@noop {} {\bibfield  {journal} {\bibinfo  {journal} {Nat. Phys.}\
}\textbf {\bibinfo {volume} {12}},\ \bibinfo {pages} {639} (\bibinfo {year}
{2016})}\BibitemShut {NoStop}%
\bibitem [{\citenamefont {Cui}\ \emph {et~al.}(2013)\citenamefont {Cui},
	\citenamefont {Lian}, \citenamefont {Ho}, \citenamefont {Lev},\ and\
	\citenamefont {Zhai}}]{Cui:2013ki}%
\BibitemOpen
\bibfield  {author} {\bibinfo {author} {\bibfnamefont {X.}~\bibnamefont
		{Cui}}, \bibinfo {author} {\bibfnamefont {B.}~\bibnamefont {Lian}}, \bibinfo
	{author} {\bibfnamefont {T.-L.}\ \bibnamefont {Ho}}, \bibinfo {author}
	{\bibfnamefont {B.~L.}\ \bibnamefont {Lev}}, \ and\ \bibinfo {author}
	{\bibfnamefont {H.}~\bibnamefont {Zhai}},\ }\href@noop {} {\bibfield
	{journal} {\bibinfo  {journal} {Phys. Rev. A}\ }\textbf {\bibinfo {volume}
		{88}},\ \bibinfo {pages} {011601} (\bibinfo {year} {2013})}\BibitemShut
{NoStop}%
\bibitem [{\citenamefont {Mancini}\ \emph {et~al.}(2015)\citenamefont
	{Mancini}, \citenamefont {Pagano}, \citenamefont {Cappellini}, \citenamefont
	{Livi}, \citenamefont {Rider}, \citenamefont {Catani}, \citenamefont {Sias},
	\citenamefont {Zoller}, \citenamefont {Inguscio}, \citenamefont {Dalmonte},\
	and\ \citenamefont {Fallani}}]{Mancini:2015fb}%
\BibitemOpen
\bibfield  {author} {\bibinfo {author} {\bibfnamefont {M.}~\bibnamefont
		{Mancini}}, \bibinfo {author} {\bibfnamefont {G.}~\bibnamefont {Pagano}},
	\bibinfo {author} {\bibfnamefont {G.}~\bibnamefont {Cappellini}}, \bibinfo
	{author} {\bibfnamefont {L.}~\bibnamefont {Livi}}, \bibinfo {author}
	{\bibfnamefont {M.}~\bibnamefont {Rider}}, \bibinfo {author} {\bibfnamefont
		{J.}~\bibnamefont {Catani}}, \bibinfo {author} {\bibfnamefont
		{C.}~\bibnamefont {Sias}}, \bibinfo {author} {\bibfnamefont {P.}~\bibnamefont
		{Zoller}}, \bibinfo {author} {\bibfnamefont {M.}~\bibnamefont {Inguscio}},
	\bibinfo {author} {\bibfnamefont {M.}~\bibnamefont {Dalmonte}}, \ and\
	\bibinfo {author} {\bibfnamefont {L.}~\bibnamefont {Fallani}},\ }\href@noop
{} {\bibfield  {journal} {\bibinfo  {journal} {Science}\ }\textbf {\bibinfo
		{volume} {349}},\ \bibinfo {pages} {1510} (\bibinfo {year}
	{2015})}\BibitemShut {NoStop}%
\bibitem [{\citenamefont {Nascimb{\`e}ne}(2013)}]{Nascimbene:2013to}%
\BibitemOpen
\bibfield  {author} {\bibinfo {author} {\bibfnamefont {S.}~\bibnamefont
		{Nascimb{\`e}ne}},\ }\href@noop {} {\bibfield  {journal} {\bibinfo  {journal}
		{J. Phys. B}\ }\textbf {\bibinfo {volume} {46}},\ \bibinfo {pages} {134005}
	(\bibinfo {year} {2013})}\BibitemShut {NoStop}%
\bibitem [{\citenamefont {Wall}\ \emph {et~al.}(2016)\citenamefont {Wall},
	\citenamefont {Koller}, \citenamefont {Li}, \citenamefont {Zhang},
	\citenamefont {Cooper}, \citenamefont {Ye},\ and\ \citenamefont
	{Rey}}]{Wall:2016cl}%
\BibitemOpen
\bibfield  {author} {\bibinfo {author} {\bibfnamefont {M.~L.}\ \bibnamefont
		{Wall}}, \bibinfo {author} {\bibfnamefont {A.~P.}\ \bibnamefont {Koller}},
	\bibinfo {author} {\bibfnamefont {S.}~\bibnamefont {Li}}, \bibinfo {author}
	{\bibfnamefont {X.}~\bibnamefont {Zhang}}, \bibinfo {author} {\bibfnamefont
		{N.~R.}\ \bibnamefont {Cooper}}, \bibinfo {author} {\bibfnamefont
		{J.}~\bibnamefont {Ye}}, \ and\ \bibinfo {author} {\bibfnamefont {A.~M.}\
		\bibnamefont {Rey}},\ }\href@noop {} {\bibfield  {journal} {\bibinfo
		{journal} {Phys. Rev. Lett.}\ }\textbf {\bibinfo {volume} {116}},\ \bibinfo
	{pages} {035301} (\bibinfo {year} {2016})}\BibitemShut {NoStop}%
\bibitem [{\citenamefont {Micheli}\ \emph {et~al.}(2010)\citenamefont
	{Micheli}, \citenamefont {Idziaszek}, \citenamefont {Pupillo}, \citenamefont
	{Baranov}, \citenamefont {Zoller},\ and\ \citenamefont
	{Julienne}}]{Micheli:2010jq}%
\BibitemOpen
\bibfield  {author} {\bibinfo {author} {\bibfnamefont {A.}~\bibnamefont
		{Micheli}}, \bibinfo {author} {\bibfnamefont {Z.}~\bibnamefont {Idziaszek}},
	\bibinfo {author} {\bibfnamefont {G.}~\bibnamefont {Pupillo}}, \bibinfo
	{author} {\bibfnamefont {M.~A.}\ \bibnamefont {Baranov}}, \bibinfo {author}
	{\bibfnamefont {P.}~\bibnamefont {Zoller}}, \ and\ \bibinfo {author}
	{\bibfnamefont {P.~S.}\ \bibnamefont {Julienne}},\ }\href@noop {} {\bibfield
	{journal} {\bibinfo  {journal} {Phys. Rev. Lett.}\ }\textbf {\bibinfo
		{volume} {105}},\ \bibinfo {pages} {073202} (\bibinfo {year}
	{2010})}\BibitemShut {NoStop}%
\bibitem [{not()}]{note}%
\BibitemOpen
\href@noop {} {}\bibinfo {note} {{Note added: After submission of the present
		work as arXiv:1607.07338 we have become aware of arXiv:1609.01285,
		Subwavelength-width optical tunnel junctions for ultracold atoms by F.
		Jendrzejewski et al., which overlaps with the first part of the present
		manuscript.}}\BibitemShut {Stop}%
\end{thebibliography}

\begin{thebibliography}{15}%
	\makeatletter
	\providecommand \@ifxundefined [1]{%
		\@ifx{#1\undefined}
	}%
	\providecommand \@ifnum [1]{%
		\ifnum #1\expandafter \@firstoftwo
		\else \expandafter \@secondoftwo
		\fi
	}%
	\providecommand \@ifx [1]{%
		\ifx #1\expandafter \@firstoftwo
		\else \expandafter \@secondoftwo
		\fi
	}%
	\providecommand \natexlab [1]{#1}%
	\providecommand \enquote  [1]{``#1''}%
	\providecommand \bibnamefont  [1]{#1}%
	\providecommand \bibfnamefont [1]{#1}%
	\providecommand \citenamefont [1]{#1}%
	\providecommand \href@noop [0]{\@secondoftwo}%
	\providecommand \href [0]{\begingroup \@sanitize@url \@href}%
	\providecommand \@href[1]{\@@startlink{#1}\@@href}%
	\providecommand \@@href[1]{\endgroup#1\@@endlink}%
	\providecommand \@sanitize@url [0]{\catcode `\\12\catcode `\$12\catcode
		`\&12\catcode `\#12\catcode `\^12\catcode `\_12\catcode `\%12\relax}%
	\providecommand \@@startlink[1]{}%
	\providecommand \@@endlink[0]{}%
	\providecommand \url  [0]{\begingroup\@sanitize@url \@url }%
	\providecommand \@url [1]{\endgroup\@href {#1}{\urlprefix }}%
	\providecommand \urlprefix  [0]{URL }%
	\providecommand \Eprint [0]{\href }%
	\providecommand \doibase [0]{http://dx.doi.org/}%
	\providecommand \selectlanguage [0]{\@gobble}%
	\providecommand \bibinfo  [0]{\@secondoftwo}%
	\providecommand \bibfield  [0]{\@secondoftwo}%
	\providecommand \translation [1]{[#1]}%
	\providecommand \BibitemOpen [0]{}%
	\providecommand \bibitemStop [0]{}%
	\providecommand \bibitemNoStop [0]{.\EOS\space}%
	\providecommand \EOS [0]{\spacefactor3000\relax}%
	\providecommand \BibitemShut  [1]{\csname bibitem#1\endcsname}%
	\let\auto@bib@innerbib\@empty
	%</preamble>
	\bibitem [{\citenamefont {Merzbacher}(1998)}]{Merzbacher1998}%
	\BibitemOpen
	\bibfield  {author} {\bibinfo {author} {\bibfnamefont {E.}~\bibnamefont
			{Merzbacher}},\ }\href@noop {} {\emph {\bibinfo {title} {Quantum
				Mechanics}}}\ (\bibinfo  {publisher} {Wiley},\ \bibinfo {year}
	{1998})\BibitemShut {NoStop}%
	\bibitem [{\citenamefont {Bloch}\ \emph {et~al.}(2008)\citenamefont {Bloch},
		\citenamefont {Dalibard},\ and\ \citenamefont {Zwerger}}]{bloch2008many}%
	\BibitemOpen
	\bibfield  {author} {\bibinfo {author} {\bibfnamefont {I.}~\bibnamefont
			{Bloch}}, \bibinfo {author} {\bibfnamefont {J.}~\bibnamefont {Dalibard}}, \
		and\ \bibinfo {author} {\bibfnamefont {W.}~\bibnamefont {Zwerger}},\
	}\href@noop {} {\bibfield  {journal} {\bibinfo  {journal} {Rev. Mod. Phys.}\
	}\textbf {\bibinfo {volume} {80}},\ \bibinfo {pages} {885} (\bibinfo {year}
	{2008})}\BibitemShut {NoStop}%
\bibitem [{\citenamefont {Yi}\ and\ \citenamefont {You}(2000)}]{You2000}%
\BibitemOpen
\bibfield  {author} {\bibinfo {author} {\bibfnamefont {S.}~\bibnamefont
		{Yi}}\ and\ \bibinfo {author} {\bibfnamefont {L.}~\bibnamefont {You}},\
}\href {\doibase 10.1103/PhysRevA.61.041604} {\bibfield  {journal} {\bibinfo
	{journal} {Phys. Rev. A}\ }\textbf {\bibinfo {volume} {61}},\ \bibinfo
{pages} {041604} (\bibinfo {year} {2000})}\BibitemShut {NoStop}%
\bibitem [{\citenamefont {Ronen}\ \emph {et~al.}(2006)\citenamefont {Ronen},
	\citenamefont {Bortolotti}, \citenamefont {Blume},\ and\ \citenamefont
	{Bohn}}]{Bohn2006}%
\BibitemOpen
\bibfield  {author} {\bibinfo {author} {\bibfnamefont {S.}~\bibnamefont
		{Ronen}}, \bibinfo {author} {\bibfnamefont {D.~C.~E.}\ \bibnamefont
		{Bortolotti}}, \bibinfo {author} {\bibfnamefont {D.}~\bibnamefont {Blume}}, \
	and\ \bibinfo {author} {\bibfnamefont {J.~L.}\ \bibnamefont {Bohn}},\ }\href
{\doibase 10.1103/PhysRevA.74.033611} {\bibfield  {journal} {\bibinfo
		{journal} {Phys. Rev. A}\ }\textbf {\bibinfo {volume} {74}},\ \bibinfo
	{pages} {033611} (\bibinfo {year} {2006})}\BibitemShut {NoStop}%
\bibitem [{\citenamefont {Sinha}\ and\ \citenamefont
	{Santos}(2007)}]{Santos2007}%
\BibitemOpen
\bibfield  {author} {\bibinfo {author} {\bibfnamefont {S.}~\bibnamefont
		{Sinha}}\ and\ \bibinfo {author} {\bibfnamefont {L.}~\bibnamefont {Santos}},\
}\href {\doibase 10.1103/PhysRevLett.99.140406} {\bibfield  {journal}
{\bibinfo  {journal} {Phys. Rev. Lett.}\ }\textbf {\bibinfo {volume} {99}},\
\bibinfo {pages} {140406} (\bibinfo {year} {2007})}\BibitemShut {NoStop}%
\bibitem [{\citenamefont {Gerbier}\ \emph {et~al.}(2005)\citenamefont
	{Gerbier}, \citenamefont {Widera}, \citenamefont {F\"olling}, \citenamefont
	{Mandel}, \citenamefont {Gericke},\ and\ \citenamefont
	{Bloch}}]{Gerbier2005}%
\BibitemOpen
\bibfield  {author} {\bibinfo {author} {\bibfnamefont {F.}~\bibnamefont
		{Gerbier}}, \bibinfo {author} {\bibfnamefont {A.}~\bibnamefont {Widera}},
	\bibinfo {author} {\bibfnamefont {S.}~\bibnamefont {F\"olling}}, \bibinfo
	{author} {\bibfnamefont {O.}~\bibnamefont {Mandel}}, \bibinfo {author}
	{\bibfnamefont {T.}~\bibnamefont {Gericke}}, \ and\ \bibinfo {author}
	{\bibfnamefont {I.}~\bibnamefont {Bloch}},\ }\href {\doibase
	10.1103/PhysRevA.72.053606} {\bibfield  {journal} {\bibinfo  {journal} {Phys.
			Rev. A}\ }\textbf {\bibinfo {volume} {72}},\ \bibinfo {pages} {053606}
	(\bibinfo {year} {2005})}\BibitemShut {NoStop}%
\bibitem [{\citenamefont {Bloch}(2005)}]{Bloch2005}%
\BibitemOpen
\bibfield  {author} {\bibinfo {author} {\bibfnamefont {I.}~\bibnamefont
		{Bloch}},\ }\href@noop {} {\bibfield  {journal} {\bibinfo  {journal} {Nature
			Physics}\ }\textbf {\bibinfo {volume} {1}},\ \bibinfo {pages} {23} (\bibinfo
	{year} {2005})}\BibitemShut {NoStop}%
\bibitem [{\citenamefont {St\"oferle}\ \emph {et~al.}(2004)\citenamefont
	{St\"oferle}, \citenamefont {Moritz}, \citenamefont {Schori}, \citenamefont
	{K\"ohl},\ and\ \citenamefont {Esslinger}}]{Stoferle2004}%
\BibitemOpen
\bibfield  {author} {\bibinfo {author} {\bibfnamefont {T.}~\bibnamefont
		{St\"oferle}}, \bibinfo {author} {\bibfnamefont {H.}~\bibnamefont {Moritz}},
	\bibinfo {author} {\bibfnamefont {C.}~\bibnamefont {Schori}}, \bibinfo
	{author} {\bibfnamefont {M.}~\bibnamefont {K\"ohl}}, \ and\ \bibinfo {author}
	{\bibfnamefont {T.}~\bibnamefont {Esslinger}},\ }\href {\doibase
	10.1103/PhysRevLett.92.130403} {\bibfield  {journal} {\bibinfo  {journal}
		{Phys. Rev. Lett.}\ }\textbf {\bibinfo {volume} {92}},\ \bibinfo {pages}
	{130403} (\bibinfo {year} {2004})}\BibitemShut {NoStop}%
\bibitem [{\citenamefont {Kollath}\ \emph {et~al.}(2006)\citenamefont
	{Kollath}, \citenamefont {Iucci}, \citenamefont {Giamarchi}, \citenamefont
	{Hofstetter},\ and\ \citenamefont {Schollw\"ock}}]{Kollath2006}%
\BibitemOpen
\bibfield  {author} {\bibinfo {author} {\bibfnamefont {C.}~\bibnamefont
		{Kollath}}, \bibinfo {author} {\bibfnamefont {A.}~\bibnamefont {Iucci}},
	\bibinfo {author} {\bibfnamefont {T.}~\bibnamefont {Giamarchi}}, \bibinfo
	{author} {\bibfnamefont {W.}~\bibnamefont {Hofstetter}}, \ and\ \bibinfo
	{author} {\bibfnamefont {U.}~\bibnamefont {Schollw\"ock}},\ }\href {\doibase
	10.1103/PhysRevLett.97.050402} {\bibfield  {journal} {\bibinfo  {journal}
		{Phys. Rev. Lett.}\ }\textbf {\bibinfo {volume} {97}},\ \bibinfo {pages}
	{050402} (\bibinfo {year} {2006})}\BibitemShut {NoStop}%
\bibitem [{\citenamefont {Gupta}\ \emph {et~al.}(2003)\citenamefont {Gupta},
	\citenamefont {Hadzibabic}, \citenamefont {Zwierlein}, \citenamefont {Stan},
	\citenamefont {Dieckmann}, \citenamefont {Schunck}, \citenamefont
	{Van~Kempen}, \citenamefont {Verhaar},\ and\ \citenamefont
	{Ketterle}}]{Gupta2003}%
\BibitemOpen
\bibfield  {author} {\bibinfo {author} {\bibfnamefont {S.}~\bibnamefont
		{Gupta}}, \bibinfo {author} {\bibfnamefont {Z.}~\bibnamefont {Hadzibabic}},
	\bibinfo {author} {\bibfnamefont {M.}~\bibnamefont {Zwierlein}}, \bibinfo
	{author} {\bibfnamefont {C.}~\bibnamefont {Stan}}, \bibinfo {author}
	{\bibfnamefont {K.}~\bibnamefont {Dieckmann}}, \bibinfo {author}
	{\bibfnamefont {C.}~\bibnamefont {Schunck}}, \bibinfo {author} {\bibfnamefont
		{E.}~\bibnamefont {Van~Kempen}}, \bibinfo {author} {\bibfnamefont
		{B.}~\bibnamefont {Verhaar}}, \ and\ \bibinfo {author} {\bibfnamefont
		{W.}~\bibnamefont {Ketterle}},\ }\href@noop {} {\bibfield  {journal}
	{\bibinfo  {journal} {Science}\ }\textbf {\bibinfo {volume} {300}},\ \bibinfo
	{pages} {1723} (\bibinfo {year} {2003})}\BibitemShut {NoStop}%
\bibitem [{\citenamefont {Shin}\ \emph {et~al.}(2007)\citenamefont {Shin},
	\citenamefont {Schunck}, \citenamefont {Schirotzek},\ and\ \citenamefont
	{Ketterle}}]{Shin2007}%
\BibitemOpen
\bibfield  {author} {\bibinfo {author} {\bibfnamefont {Y.}~\bibnamefont
		{Shin}}, \bibinfo {author} {\bibfnamefont {C.~H.}\ \bibnamefont {Schunck}},
	\bibinfo {author} {\bibfnamefont {A.}~\bibnamefont {Schirotzek}}, \ and\
	\bibinfo {author} {\bibfnamefont {W.}~\bibnamefont {Ketterle}},\ }\href
{\doibase 10.1103/PhysRevLett.99.090403} {\bibfield  {journal} {\bibinfo
		{journal} {Phys. Rev. Lett.}\ }\textbf {\bibinfo {volume} {99}},\ \bibinfo
	{pages} {090403} (\bibinfo {year} {2007})}\BibitemShut {NoStop}%
\bibitem [{\citenamefont {T\"orm\"a}\ and\ \citenamefont
	{Zoller}(2000)}]{Torma2000}%
\BibitemOpen
\bibfield  {author} {\bibinfo {author} {\bibfnamefont {P.}~\bibnamefont
		{T\"orm\"a}}\ and\ \bibinfo {author} {\bibfnamefont {P.}~\bibnamefont
		{Zoller}},\ }\href {\doibase 10.1103/PhysRevLett.85.487} {\bibfield
	{journal} {\bibinfo  {journal} {Phys. Rev. Lett.}\ }\textbf {\bibinfo
		{volume} {85}},\ \bibinfo {pages} {487} (\bibinfo {year} {2000})}\BibitemShut
{NoStop}%
\bibitem [{\citenamefont {Chin}\ \emph {et~al.}(2004)\citenamefont {Chin},
	\citenamefont {Bartenstein}, \citenamefont {Altmeyer}, \citenamefont {Riedl},
	\citenamefont {Jochim}, \citenamefont {Denschlag},\ and\ \citenamefont
	{Grimm}}]{Chin2004}%
\BibitemOpen
\bibfield  {author} {\bibinfo {author} {\bibfnamefont {C.}~\bibnamefont
		{Chin}}, \bibinfo {author} {\bibfnamefont {M.}~\bibnamefont {Bartenstein}},
	\bibinfo {author} {\bibfnamefont {A.}~\bibnamefont {Altmeyer}}, \bibinfo
	{author} {\bibfnamefont {S.}~\bibnamefont {Riedl}}, \bibinfo {author}
	{\bibfnamefont {S.}~\bibnamefont {Jochim}}, \bibinfo {author} {\bibfnamefont
		{J.~H.}\ \bibnamefont {Denschlag}}, \ and\ \bibinfo {author} {\bibfnamefont
		{R.}~\bibnamefont {Grimm}},\ }\href@noop {} {\bibfield  {journal} {\bibinfo
		{journal} {Science}\ }\textbf {\bibinfo {volume} {305}},\ \bibinfo {pages}
	{1128} (\bibinfo {year} {2004})}\BibitemShut {NoStop}%
\bibitem [{\citenamefont {Regal}\ \emph {et~al.}(2003)\citenamefont {Regal},
	\citenamefont {Ticknor}, \citenamefont {Bohn},\ and\ \citenamefont
	{Jin}}]{Regal2003}%
\BibitemOpen
\bibfield  {author} {\bibinfo {author} {\bibfnamefont {C.~A.}\ \bibnamefont
		{Regal}}, \bibinfo {author} {\bibfnamefont {C.}~\bibnamefont {Ticknor}},
	\bibinfo {author} {\bibfnamefont {J.~L.}\ \bibnamefont {Bohn}}, \ and\
	\bibinfo {author} {\bibfnamefont {D.~S.}\ \bibnamefont {Jin}},\ }\href@noop
{} {\bibfield  {journal} {\bibinfo  {journal} {Nature}\ }\textbf {\bibinfo
		{volume} {424}},\ \bibinfo {pages} {47} (\bibinfo {year} {2003})}\BibitemShut
{NoStop}%
\bibitem [{\citenamefont {\L\k{a}cki}\ \emph {et~al.}()\citenamefont
	{\L\k{a}cki}, \citenamefont {Elben}, \citenamefont {Pichler}, \citenamefont
	{Baranov},\ and\ \citenamefont {Zoller}}]{Lacki2016inp}%
\BibitemOpen
\bibfield  {author} {\bibinfo {author} {\bibfnamefont {M.}~\bibnamefont
		{\L\k{a}cki}}, \bibinfo {author} {\bibfnamefont {A.}~\bibnamefont {Elben}},
	\bibinfo {author} {\bibfnamefont {H.}~\bibnamefont {Pichler}}, \bibinfo
	{author} {\bibfnamefont {M.}~\bibnamefont {Baranov}}, \ and\ \bibinfo
	{author} {\bibfnamefont {P.}~\bibnamefont {Zoller}},\ }\href@noop {}
{\bibinfo  {journal} {in preparation}\ }\BibitemShut {NoStop}%
\end{thebibliography}
\end{document}